\documentclass[12pt]{JHEP3}

\usepackage{amsmath,amssymb}
\usepackage{latexsym}
\usepackage{graphicx}
\usepackage{epsfig}

\newcommand{\beqs}{\begin{equation*}}
\newcommand{\beq}{\begin{equation}}

\newcommand{\eeqs}{\end{equation*}}
\newcommand{\eeq}{\end{equation}}

\newcommand{\beqas}{\begin{eqnarray*}}
\newcommand{\beqa}{\begin{eqnarray}}

\newcommand{\eeqas}{\end{eqnarray*}}
\newcommand{\eeqa}{\end{eqnarray}}

\newcommand{\eq}[2]{\begin{equation} #1 \label{#2} \end{equation}}

\newcommand{\eps}{\varepsilon}
\newcommand{\al}{\alpha}

\newcommand{\de}{\delta}
\newcommand{\om}{\omega}

\newcommand{\Ga}{\Gamma}

\newcommand{\Om}{\Omega}

\newcommand{\blist}{\begin{itemize}}

\newcommand{\elist}{\end{itemize}}

\providecommand{\href}[2]{#2}

\DeclareFontFamily{OT1}{rsfs}{}
\DeclareFontShape{OT1}{rsfs}{m}{n}{ <-7> rsfs5 <7-10> rsfs7 <10->rsfs10}{} 
\DeclareMathAlphabet{\mycal}{OT1}{rsfs}{m}{n}

\newcommand{\gthree}{{\gamma_\ast}}

\newcommand{\jq}{{j_q}}
\newcommand{\jp}{{j_p}}

\newcommand{\medsp}{\\[0.7ex]}

\newcommand{\matp}{\mathfrak{p}}
\newcommand{\matq}{\mathfrak{q}}

\newcommand{\ve}{\varepsilon}

\newcommand{\dega}{\ensuremath{^\dag}}

\newcommand{\diff}[1][]{\mbox{d}#1}

\newcommand{\gf}[1]{\itindex{#1}{gf}}
\newcommand{\half}[1]{\ensuremath{\frac{#1}{2}}}
\newcommand{\intd}[1]{\int \!\! #1 \;}
\newcommand{\inv}[1]{\ensuremath{\frac{1}{#1}}}

\newcommand{\Stext}[1]{\itindex{\mathcal{S}}{#1}}

\newcommand{\derfrac}[2][]{\frac{\partial #1}{\partial #2}}

\newcommand{\itindex}[2]{\ensuremath{#1_{\mbox{\scriptsize{\itshape #2}}}}}

\newcommand{\varfrac}[2][]{\frac{\delta #1}{\delta #2}}

\DeclareMathOperator{\extdm}{d}
\newcommand{\extd}{\extdm \!}

\title{Quantization of 2D dilaton supergravity with matter}

\author{L. Bergamin\footnotemark[1],
  D. Grumilller\footnotemark[1] \footnotemark[2] and
  W. Kummer\footnotemark[1]\\ \footnotemark[1]\,\,\parbox[t]{13cm}{Institute
  for Theoretical Physics, Vienna University
  of Technology\\ 
  Wiedner Hauptstra{\ss}e 8-10, A-1040 Vienna, Austria 
  }\\ \ \\ 
  \footnotemark[2]\,\,\parbox[t]{13cm}{Institute for Theoretical Physics, Leipzig University
    \\
    Augustusplatz 10-11, D-04103 Leipzig, Germany
  }\\ \ \\ Email: \email{bergamin@tph.tuwien.ac.at}, \email{grumil@hep.itp.tuwien.ac.at},
  \email{wkummer@tph.tuwien.ac.at}.
}

\abstract{General $N=(1,1)$ dilaton supergravity in two dimensions allows a 
background independent exact quantization of the geometric part, 
if these theories are formulated as  specific graded Poisson-sigma 
models. The strategy developed for the bosonic case can be carried 
over, although considerable computational complications arise 
when the Hamiltonian constraints are evaluated in the presence of 
matter. Nevertheless, the constraint structure is the same as in the bosonic theory. In the matterless case gauge independent nonlocal correlators are
calculated non-perturbatively. They respect local quantum triviality and allow
a topological interpretation.

In the presence of matter the ensuing nonlocal effective theory is expanded in
matter loops. The lowest order tree vertices are derived and discussed,
entailing the phenomenon of virtual black holes which essentially determine
the corresponding S-matrix. Not all vertices are conformally invariant, but
the S-matrix is invariant, as expected.

Finally, the proper measure for the 1-loop corrections is addressed. It is
argued how to exploit the results from fixed background quantization for our
purposes.}

\preprint{TUW-04-04\\LU-ITP-04-004}


\begin{document}
\numberwithin{equation}{section}

\section{Introduction}
The traditional formulation of $N=(1,1)$ dilaton supergravity 
models in two dimensions is based upon superfields \cite{Howe:1979ia,Park:1993sd}. In 
this approach the bosonic part of the supertorsion is set to zero and the fermionic component is assumed to 
be the one of flat supergravity \cite{Howe:1979ia}. In this way the 
Bianchi identities do not pose any problem, because they turn out 
to be fulfilled identically. Also, it is convenient to fix certain components
in the superzweibein by a suitable gauge choice.

Although quite a number of interesting results has been obtained 
in this way, the complications due to the large number of 
auxiliary fields in the past have limited  applications mostly to 
statements about the bosonic part of such theories for which ref.\
\cite{Park:1993sd} is a key reference.

In the last years an alternative treatment of general 
\emph{bosonic} dilaton theories has been very successful. It is 
based upon the local and global equivalence \cite{Katanaev:1996bh} of such 
theories to so-called Poisson-sigma models (PSMs) \cite{Schaller:1994es} of 
gravity where beside the zweibein also the spin 
connection appears as an \emph{independent} variable 
together with new scalar fields (``target-space coordinates'') 
on the 2D world sheet. Although those formulations possess 
non-vanishing bosonic torsion and thus a larger configuration 
space, the classical, as well as the quantum treatment, have been 
found to be much simpler. Based upon earlier work \cite{Kummer:1992rt,Haider:1994cw} which had 
observed these features already before the advent of the PSM, many 
novel results could be obtained which seem to be inaccessible in 
practice in the usual dilaton field approach. They include the 
exact (background independent) quantization of the geometric part 
\cite{Haider:1994cw,Kummer:1997hy,Kummer:1998zs}, the confirmation of the existence of ``virtual 
Black Holes'' without any further ad hoc assumptions \cite{Grumiller:2000ah,Grumiller:2002dm}, the 
computation of the one-loop correction to the specific heat 
\cite{Grumiller:2003mc,Zaslavsky:1996dg} in the string inspired
\cite{Mandal:1991tz} dilaton Black Hole model \cite{Callan:1992rs}
etc.\footnote{A comprehensive review putting the results of the 
PSM-approach into the perspective of previous work on 2D gravity 
theories is \cite{Grumiller:2002nm}. Actually all known models (Jackiw-Teitelboim 
\cite{Barbashov:1979bm}, spherically reduced Einstein gravity \cite{Thomi:1984na}, the CGHS 
model \cite{Callan:1992rs} etc.) are special cases.} It should be pointed out
that PSMs have attracted attention in string theory \cite{Schomerus:1999ug}
due to the path integral interpretation \cite{Cattaneo:1999fm} of Kontsevich's
$\star$-product \cite{Kontsevich:1997vb}.

In order to benefit from these promising features also in the 
context of supergravity \cite{Freedman:1976xh} it appeared natural to
exploit \emph{graded} Poisson-sigma models (gPSMs), where the 
target space is extended by a ``dilatino'' and the gauge fields 
comprise also the gravitino \cite{Izquierdo:1998hg,Ertl:2000si}. Recently two of the 
present authors (L.B.\ and W.K.) succeeded \cite{Bergamin:2002ju,Bergamin:2003am} in 
identifying a certain subclass of gPSMs with the 
``genuine'' dilaton supergravity of ref.\ \cite{Park:1993sd} when certain 
components of the superfield in the latter are properly 
expressed in terms of the fields appearing in the gPSM formulation. 
In the parlance of refs.\  \cite{Bergamin:2002ju,Bergamin:2003am} the latter is called ``minimal 
field supergravity'' (MFS) in the following. MFS yielded the 
first complete solution (including fermions) for the 2D supergravity 
of ref.\ \cite{Park:1993sd}, the proper formulation of the super-pointparticle in the 
background of such a solution (and thus first nontrivial  
examples of ``supergeodesics''), but also a full analysis of 
solutions retaining some supersymmetries, like BPS-states \cite{Bergamin:2003mh}. 
In the latter paper also minimal and non-minimal interactions with matter
could be introduced at the gPSM level.

Spurred by the successful quantization in the PSM of bosonic 
gravity, in our present paper we carry out an analogous program 
for $N=(1,1)$ supergravity, minimally coupled 
to matter. This comprises as an essential intermediate step the 
evaluation of the constraint algebra. The strategy of 
bosonic theories  can be carried over \cite{Haider:1994cw,Kummer:1997hy,Kummer:1998zs}: Geometry is 
integrated out first in a background-independent way. Certain 
boundary conditions for the geometry are fixed which selects a 
suitable background a posteriori. Finally, the effective 
theory is expanded perturbatively in matter loops, taking fully 
into account backreactions to each order in a self-consistent 
way.

This paper is organized as follows: section \ref{sec:two} briefly reviews the
relations between gPSMs and MFS and the coupling to
matter fields. In section \ref{sec:constraints} a Hamiltonian analysis is
performed. This is the necessary prerequisite for the construction of the BRST
charge and the path integral quantization in section \ref{sec:3} employing a
suitable gauge-fixing fermion. In the absence of matter the theory is found to
be locally quantum trivial, although nontrivial nonlocal correlators
exist. The lowest order matter vertices are derived in section
\ref{sec:vertices} and their implications for the S-matrix are
addressed. Section \ref{sec:loops} is devoted to 1-loop corrections. The
conclusions in \ref{sec:conclusions} contain comments on generalizations to
non-minimal coupling and to local self-interactions. Appendix
\ref{sec:notation} summarizes our notations and conventions. Some details
which are relevant for the path integral, but which are somewhat decoupled
from the main text, namely the treatment of boundary terms and the proof
regarding the absence of ordering problems, can be found in appendix \ref{app:B}.

\section{Graded Poisson-Sigma model and minimal field supergravity}
\label{sec:two}
\subsection{Graded Poisson-Sigma model}
\label{sec:twoone}
A general gPSM consists of scalar fields
$X^I(x)$, which are themselves (``target space'') coordinates of a graded Poisson manifold with
Poisson tensor $P^{IJ}(X) = (-1)^{IJ+1} P^{JI}(X)$. The index
$I$, in the generic case, may include commuting as well as anti-commuting
fields\footnote{Details on the usage of different indices as well as other features of
  our notation are given in
  Appendix \ref{sec:notation}.
For further details one should consult ref.\ \cite{Ertl:2000si,Ertl:2001sj}.}. In addition one introduces the gauge
potential $A = \extd X^I A_I = \extd X^I A_{mI}(x) \extd x^m$, a one form with respect to the Poisson
structure as well as with respect to the 2D worldsheet. The gPSM
action reads\footnote{If the multiplication of forms is evident the wedge symbol will be omitted subsequently.}
\begin{equation}
  \label{eq:gPSMaction}
  \begin{split}
    \Stext{gPSM} &= \int_{\mathcal{M}} \extd X^I \wedge A_I + \half{1} P^{IJ}
    A_J \wedge A_I\ . 
  \end{split}  
\end{equation}
The Poisson tensor $P^{IJ}$ must have vanishing Nijenhuis tensor (obeying a
Jacobi-type identity with respect to the Schouten bracket related to the Poisson tensor as $\{ X^I,X^J \} = P^{IJ}$)
\begin{equation}
\label{eq:nijenhuis}
  P^{IL}\partial _{L}P^{JK}+ \mbox{\it 
g-perm}\left( IJK\right) = 0\ ,
\end{equation}
where the sum runs over the graded permutations. Due to
\eqref{eq:nijenhuis} the action \eqref{eq:gPSMaction} is invariant under the
symmetry transformations
\begin{align}
\label{eq:symtrans}
  \delta X^{I} &= P^{IJ} \ve _{J}\ , & \delta A_{I} &= -\mbox{d} \ve
  _{I}-\left( \partial _{I}P^{JK}\right) \ve _{K}\, A_{J}\ ,
\end{align}
where the term $\extd \epsilon_I$ in the second of these equations provides
the justification for calling $A_I$ ``gauge fields''.

For a generic (g)PSM the commutator of two transformations \eqref{eq:symtrans} is a
symmetry modulo the equations of motion. Only for \( P^{IJ} \) linear
in \( X^{I} \) a closed (and linear) algebra is obtained, namely a graded Lie algebra; in this case
\eqref{eq:nijenhuis} reduces to the Jacobi identity for the structure
constants thereof. If the Poisson
tensor has a non-vanishing kernel---the actual situation in any application to 2D
(super-)gravity due to the odd dimension of the bosonic part of the tensor---there exist (one or more) Casimir functions $C(X)$ obeying
\begin{equation}
\label{eq:casimir}
  \{ X^I, C \} = P^{IJ}\derfrac[C]{X^J} = 0\ ,
\end{equation}
which, when determined by the field equations
\begin{align}
\label{eq:gPSMeom1}
  \extd X^I + P^{IJ} A_J &= 0\ ,\medsp
\label{eq:gPSMeom2}
  \extd A_I + \half{1} (\partial_I P^{JK}) A_K A_J &= 0\ ,
\end{align}
are constants of
motion.

In the most immediate application to 2D supergravity\footnote{More complicated identifications of the 2D Cartan variables with
\( A_{I} \) are conceivable \cite{Strobl:2003kb}.} the gauge
potentials comprise the spin connection $\om^a{}_b=\eps^a{}_b\om$, the dual basis $e_a$ containing the zweibein and the gravitino $\psi_\al$:
\begin{align}
  A_I &= (A_\phi, A_a, A_\alpha) = (\omega, e_a, \psi_\alpha) & X^I &= (X^\phi,
  X^a, X^\alpha) = (\phi, X^a, \chi^\alpha)
\end{align}
The fermionic components \( \psi _{\alpha } \) (``gravitino'') and \( \chi
^{\alpha } \) (``dilatino'') for $N=(1,1)$ supergravity are Majorana spinors. The scalar field $\phi$ will be referred to as ``dilaton''. The remaining bosonic target space coordinates $X^a$ correspond to directional derivatives of the dilaton in the second order formulation presented below (cf.\ eq.\ \ref{eq:GDT}).
Local Lorentz invariance determines the $\phi$-components of the Poisson
tensor
\begin{align}
\label{eq:lorentzcov}
  P^{a \phi} &= X^b {\epsilon_b}^a\ , & P^{\alpha \phi} &= -\half{1}
  \chi^\beta {\gthree_\beta}^\alpha\ ,
\end{align}
and the supersymmetry transformation is encoded in $P^{\alpha \beta}$.

\subsection{Recapitulation of dilaton gravity}

In a purely bosonic theory, the only non-trivial component of the
Poisson tensor is $P^{ab} = v \epsilon^{ab}$, where the locally Lorentz invariant 
``potential'' \( v=v\left( \phi ,Y\right)  \) describes different
models (\( Y=X^{a}X_{a}/2 \)). Evaluating \eqref{eq:gPSMaction} with that
$P^{ab}$ and $P^{a \phi}$ from \eqref{eq:lorentzcov}, the action
\cite{Schaller:1994es}
\begin{equation}
  \label{eq:bosonicPSM}
  \Stext{PSM} = \int_{\mathcal{M}} \bigl( \phi \diff \omega + X^a D e_a +
  \epsilon v \bigr)
\end{equation}
is obtained. $\epsilon =
\half{1} \epsilon^{ab} e_b \wedge e_a$ is the volume form and the covariant
$D$ in the torsion term is defined in (\ref{eq:A8}). The most
interesting models are described by potentials quadratic in $X^a$
\begin{equation}
  \label{eq:bosonpot}
  v=Y\, Z\left( \phi \right) +V\left( \phi \right)\ .
\end{equation}
All physically interesting
2D dilaton theories (spherically reduced Einstein gravity
\cite{Thomi:1984na}, the string
inspired
black hole \cite{Callan:1992rs}, the Jackiw-Teitelboim model with \( Z=0 \) and
linear \( V\left( \phi \right)  \) \cite{Barbashov:1979bm},
the bosonic part of the Howe model \cite{Howe:1979ia} etc.)\ are expressible
by a potential of type \eqref{eq:bosonpot}. They allow the integration of the
(single) Casimir function
$C$ in \eqref{eq:casimir}
\begin{align}
\label{eq:bosonicC}
  C&= e^{Q(\phi)} Y + W(\phi)\ , & Q(\phi) &= \int_{\phi_1}^\phi \extd \varphi
  Z(\varphi)\ , & W(\phi) &= \int_{\phi_0}^\phi \extd \varphi e^{Q(\varphi)}
  V(\varphi)\ ,
\end{align}
where e.g. in spherically reduced gravity $C$ on-shell is proportional
to the ADM-mass in the Schwarzschild solution. Its conservation $\extd C=0$
had been found previously in refs.~\cite{Banks:1991mk}.

The auxiliary variables
\( X^{a} \) and the torsion-dependent part of the spin connection $\omega$
can be eliminated as they appear linear in the relevant equations of motion. Then the action
reduces to the familiar generalized dilaton theory
in terms of the dilaton field \( \phi  \) and the metric:
\begin{equation}
\label{eq:GDT}
  \Stext{GDT} = \intd{\diff{^2 x}} \sqrt{-g} \Bigl(\half{1} R \phi - \half{1} Z
  \partial^m \phi \partial_m \phi + V(\phi) \Bigr)
\end{equation}
Both formulations are equivalent at the classical \cite{Katanaev:1996bh,Katanaev:1997ni} as
well as at the quantum level
\cite{Kummer:1997hy,Kummer:1998zs}.

For theories with non-dynamical dilaton ($Z = 0$ in \eqref{eq:bosonpot}) a further
elimination of $\phi$ is possible if the potential $V(\phi)$ is
invertible. In this way one  arrives at a theory solely formulated in terms of
the metric $g_{mn}=e^a_me^b_n\eta_{ab}$ \cite{Obukhov:1997uc}.
\subsection{Minimal field supergravity}
\label{sec:twotwo}
A generic fermionic extension of the action \eqref{eq:bosonicPSM} is obtained
by the most general choice of $P^{a \alpha}$, $P^{\alpha \beta}$ and the
fermionic extension of $P^{ab} = \epsilon^{ab}(v + \chi^2 v_2)$ solving
\eqref{eq:nijenhuis}. Here \eqref{eq:lorentzcov} and the bosonic potential $v$
are a given input. This leads to an algebraic, albeit highly ambiguous solution of (\ref{eq:nijenhuis}) with several arbitrary functions \cite{Ertl:2000si}. In addition, the
fermionic extensions generically exhibit new singular terms. Also
not all bosonic models permit such an extension for the whole range
of their bosonic fields; sometimes even no extension is allowed. 

As shown by two of the present authors \cite{Bergamin:2002ju}, it is, nevertheless,
possible to select {}``genuine'' supergravity from this huge set
of theories. This has been achieved by a generalization of the standard
requirements for a ``true'' supergravity
\cite{Freedman:1976xh}
to the situation, where deformations from the dilaton field $\phi$ are present. To this end the
non-linear symmetry \eqref{eq:symtrans}, which is closed on-shell only, is---in a first step---related to the more convenient (off-shell closed) algebra of
Hamiltonian constraints $G^I = \partial_1 X^I + P^{IJ}(X) A_{1 J}$ discussed
in detail in Section \ref{sec:constraints}. The Hamiltonian obtained from
\eqref{eq:gPSMaction} is a linear combination of these constraints
\cite{Grosse:1992vc,Haider:1994cw,Bergamin:2002ju}:
\begin{equation}
H=\intd{\diff{x^{1}}}G^{I}A_{0I}
\end{equation}
In a second step a certain linear combination of the \( G^{I} \), suggested by the
ADM parametrization \cite{Katanaev:1994qf,Katanaev:2000kc,Bergamin:2002ju}, maps the \( G^{I} \)
algebra upon a deformed version of the superconformal algebra (deformed
Neveu-Schwarz, resp.\  Ramond algebra). This algebra is
appropriate to impose restrictions, which represent a natural
generalization of the requirements from supergravity to theories deformed by the
dilaton field. Translated into analytic restrictions onto the Poisson tensor
they take the simple form\footnote{Here and in what follows light-cone
  coordinates are used, cf.\ eqs.\ \eqref{eq:Achi}-\eqref{eq:gammalc}.}
\begin{align}
\label{eq:sugraconstr1}
  \derfrac{X^{++}}(P^{+|-}, P^{+|++}, P^{+| --}) &= 0\ , &\derfrac{X^{--}} (
  P^{+|-}, P^{-|--},  P^{-| ++}) &= 0\ , \medsp
\label{eq:sugraconstr2}
   \partial_{++} P^{++|--} - 2 \partial_+ P^{+| --}  &= 0\ , & \partial_{++}
  P^{++|-} - 2 \partial_+ P^{+|-}
  &=0\ ,
\end{align}
whereby \eqref{eq:sugraconstr2} can be derived from
\eqref{eq:sugraconstr1} together with \eqref{eq:nijenhuis}.

It turned out that the subset of
models allowed by these restrictions uniquely leads to the gPSM \emph{supergravity}
class of theories (called ``minimal field supergravity'', MFS, in our present paper) with the Poisson
tensor\footnote{The constant $\tilde{u}_0$ in ref.\ \cite{Bergamin:2002ju} has
  been fixed as
  $\tilde{u}_0 = -2$. This is in agreement with standard
  supersymmetry conventions. $f'$ denotes $d f /d \phi$.} (\( \chi ^{2}=\chi ^{\alpha }\chi _{\alpha }\), $Y = X^a X_a / 2$)
\begin{align}
\label{eq:mostgensup}
  P^{ab} &= \biggl( V + Y Z - \half{1} \chi^2 \Bigl( \frac{VZ + V'}{2u} +
  \frac{2 V^2}{u^3} \Bigr) \biggr) \epsilon^{ab}\ , \medsp
  P^{\alpha b} &= \frac{Z}{4} X^a
    {(\chi \gamma_a \gamma^b \gthree)}^\alpha + \frac{i V}{u}
  (\chi \gamma^b)^\alpha\ , \medsp
\label{eq:mostgensuplast}
  P^{\alpha \beta} &= -2 i X^c \gamma_c^{\alpha \beta} + \bigl( u +
  \frac{Z}{8} \chi^2 \bigr) \gthree^{\alpha \beta}\ ,
\end{align}
where the three functions $V$, $Z$ and the
``prepotential'' $u$ depend on the
dilaton field $\phi$ only. Besides the components of $P^{IJ}$  already fixed according to
\eqref{eq:lorentzcov}, supergravity requires the existence of supersymmetry
transformations, which are generated by the first
term in \eqref{eq:mostgensuplast}. It is a central result of ref.\
\cite{Bergamin:2002ju} that $P^{\alpha \beta}$ must be of the form
\eqref{eq:mostgensuplast}: The first term, which is not allowed to receive any deformations, is dictated by the flat limit of
rigid supersymmetry. Furthermore, in order to satisfy the condition
\eqref{eq:nijenhuis} $V$, $Z$ and $u$ must be related by
\begin{equation}
\label{eq:finalpot}
V\left( \phi \right) =- \inv{8} \bigl(( u^{2})' + u^{2} Z\left( \phi
\right) \bigr) \ .
\end{equation}
Thus, starting from a certain bosonic model with potential
\begin{equation}
v = V+YZ
\end{equation}
in \eqref{eq:mostgensup}, the only restriction is that it must be expressible in
terms of a prepotential \( u \) by \eqref{eq:finalpot}. This happens to be the
case for all physically interesting theories
\cite{Thomi:1984na,Callan:1992rs,Barbashov:1979bm,Howe:1979ia}. Inserting the Poisson tensor \eqref{eq:lorentzcov},
\eqref{eq:mostgensup}-\eqref{eq:mostgensuplast}
into equation \eqref{eq:gPSMaction} the ensuing action becomes (cf.\ eq.\
\eqref{eq:A8} for the definition of the respective covariant derivatives $D$)
\begin{multline}
  \label{eq:mostgenaction}
  \Stext{MFS} = \int_{\mathcal{M}} \bigl( \phi \diff \omega + X^a D e_a + \chi^\alpha D
  \psi_\alpha + \epsilon \biggl( V + Y Z - \half{1} \chi^2 \Bigl( \frac{VZ + V'}{2u} +
  \frac{2 V^2}{u^3} \Bigr) \biggr) \medsp
   + \frac{Z}{4} X^a
    (\chi \gamma_a \gamma^b e_b \gthree \psi) + \frac{i V}{u}
  (\chi \gamma^a e_a \psi) 
   + i X^a (\psi \gamma_a \psi) - \half{1} \bigl( u +
  \frac{Z}{8} \chi^2 \bigr) (\psi \gthree \psi) \bigr)\ .
\end{multline}
As in case of the purely bosonic theory \eqref{eq:bosonicPSM} the Poisson tensor
has at least one (bosonic) Casimir function \eqref{eq:bosonicC}. In terms of the prepotential $u$
it reads
\begin{align}
\label{eq:C1}
  C &= e^Q (X^{++} X^{--} - \inv{8} u^2 + \inv{8} \chi^- \chi^+ C_\chi)\ ,
  \medsp
\label{eq:C2}
  C_\chi &= u' + \half{1} u Z = - 4 \frac{V}{u}\ .
\end{align}
If $C = 0$ the symmetric part of the Poisson tensor is degenerate and 
a second (fermionic) Casimir function $\tilde{c}$ emerges (cf.\
\cite{Ertl:2000si,Bergamin:2003am}). 

It has been proven in \cite{Bergamin:2003am} that this class of supergravity
models is equivalent to the superfield supergravity of Park and Strominger
\cite{Park:1993sd} upon elimination of auxiliary fields on both sides, when a
certain linear combination of the gPSM gravitino and dilatino in MFS is
identified with the gravitino of superfield supergravity (cf.\ sect.\ (5.2) of
ref.\ \cite{Bergamin:2003am} and footnote \ref{foot321} below). The
classical aspects of these models have been studied in some detail in refs.\ \cite{Bergamin:2003am,Bergamin:2003mh}.
\subsection{Coupling of matter fields}
This equivalence can be used to derive from the superspace construction the
matter coupling for MFS models. For details of the calculations we
refer to \cite{Bergamin:2003mh}. A supersymmetric matter multiplet consists of
a real scalar field $f$ and a Majorana spinor $\lambda_\alpha$. In case of
non-minimal coupling a coupling function $P(\phi)$ is introduced as well. Given the technical difficulties we restrict to minimal
coupling ($P(\phi) \equiv 1$ in the notation of \cite{Bergamin:2003mh}) in the explicit calculations and
generalizations will be commented upon the end. Then the matter action
\begin{multline}
\label{eq:cmMFS}
      \mathcal{S}_{(m)} = \int_{\mathcal{M}} \Bigl( \half{1} \extd {f} \wedge
    \ast \extd {f} + \half{i}
    {\lambda} \gamma_a {e}^a \wedge \ast \extd {\lambda}\medsp + i \ast({e}_a \wedge \ast \extd {f}) {e}_b \wedge \ast {\psi} \gamma^a
    \gamma^b {\lambda}
    + \inv{4} \ast ({e}_b \wedge \ast {\psi}) \gamma^a \gamma^b
    {e}_a \wedge \ast {\psi} {\lambda}^2 \Bigr)
\end{multline}
is found to be invariant under local supersymmetry transformations.

\section{Hamiltonian analysis}
\label{sec:constraints}
The primary goal of the present paper is to develop the systematics of the
quantization of the action
\eqref{eq:mostgenaction} together with matter couplings \eqref{eq:cmMFS}.
The quantization is performed via a Hamiltonian analysis introducing Poisson
brackets. Though the appearance of fermions in
both, geometry and matter part, lead to some technical complications the final
result is seen to retain the structure already found in the bosonic
case
\cite{Grosse:1992vc,Haider:1994cw,Kummer:1997hy,Kummer:1997jr}.
For the purely geometrical part of the action the result of this analysis has
been presented  already in \cite{Bergamin:2002ju}. Nevertheless, a detailed
formulation is given.
For convenience many formulas are directly written in light-cone basis set out
in Appendix \ref{sec:notation}.

Although we will proceed by analyzing MFS together with matter, an
important remark about the range of Poisson tensors covered by our results
should be made: The analysis of matterless (super-)geometry is valid for \emph{any}
graded Poisson tensor with local Lorentz invariance, i.e.\ whose components
$P^{a\phi}$ and $P^{\alpha \phi}$ are determined by
\eqref{eq:lorentzcov}. Thus theories are covered as well that are not
matterless supergravity in the sense of ref.\ \cite{Bergamin:2002ju}. The coupling of conformal
matter fields according to the last section, however, is restricted only to
the models of ref.\ \cite{Bergamin:2002ju} (i.e.\ the actions
\eqref{eq:mostgenaction}). In what follows quantities evaluated from the gPSM part of the
action are indicated by $(g)$, while $(m)$ labels the quantities from the
matter action. This separation is possible as the matter action
\eqref{eq:cmMFS} does not contain derivatives acting on the MFS fields.

\subsection{First class constraints}

In the geometrical sector we define the canonical variables\footnote{The somewhat unusual
  association of gauge fields as ``momenta''
and of target space coordinate fields as ``coordinates'' is motivated in
  appendix \ref{sec:3.1}, where boundary conditions are discussed.} and the first class primary
constraints ($\approx$ means zero on the surface of constraints) from the Lagrangian $L_{(g)}$ in \eqref{eq:gPSMaction}
($\dot{q}^I = \partial_0 X^I$) by
\begin{align}
  X^I &= q^I\ , & \bar{q}^I &= (-1)^{I+1} \derfrac[L]{\dot{\bar{p}}_I} \approx 0\ , \medsp
\label{eq:gaugepot}
   \derfrac[L]{\dot{q}^I} &= \derfrac[L_{(g)}]{\dot{q}^I} = p_I = A_{1 I}\ , & \bar{p}_I &= A_{0 I}\ .
\end{align}
From the Hamiltonian density ($\partial_1 = \partial$)
\begin{equation}
\label{eq:hamdensity}
  H_{(g)} =  \dot{q}^I p_I - L_{(g)} = \partial q^I \bar{p}_I - P^{IJ} \bar{p}_J p_I
\end{equation}
the graded canonical equations
\begin{align}
  \derfrac[H_{(g)}]{p_I} &= (-1)^I \dot{q}^I\ , & \derfrac[H_{(g)}]{q^I} &= -
  \dot{p}_I\ ,
\end{align}
are consistent with the graded Poisson\footnote{This is the standard Poisson
  bracket, not to be confused with the Schouten bracket, associated to the
  Poisson tensor $P^{IJ}$ in eqs.\ \eqref{eq:nijenhuis}, \eqref{eq:symtrans}.}
  bracket for functionals $A$ and $B$
\begin{equation}
\label{eq:canonicalbr}
  \begin{split}
    \{A, B'\} &= \int_{x''} \Bigl[\bigl( (-1)^{A \cdot I} \varfrac[A]{{q''}^I}
    \varfrac[B']{p''_I} - (-1)^{I(A+1)} \varfrac[A]{p''_I}
    \varfrac[B']{{q''}^I}\bigr) + (q\rightarrow \bar{q}, p \rightarrow
    \bar{p}) \Bigr]\medsp
    &= \int_{x''} \biggl[ \Bigl(\bigl( \varfrac[A]{{q''}^i}
    \varfrac[B']{p''_i} - \varfrac[A]{p''_i}
    \varfrac[B']{{q''}^i}\bigl) + (-1)^A \bigl(  \varfrac[A]{{q''}^\alpha}
    \varfrac[B']{p''_\alpha} + \varfrac[A]{p''_\alpha}
    \varfrac[B']{{q''}^\alpha}\bigr)\Bigr)\medsp
    &\phantom{= \int_{x''} \biggl[ \Bigl(\bigl(}+ (q\rightarrow \bar{q}, p \rightarrow
    \bar{p}) \biggr]\ ,
  \end{split}
\end{equation}
where $(q\rightarrow \bar{q}, p \rightarrow \bar{p})$ indicates that the
functional derivatives have to be performed for both types of variables, with and without bar.
The primes indicate the dependence on primed world-sheet
coordinates $x$, resp.\ $x'$, $x''$. The Hamiltonian density \eqref{eq:hamdensity}
\begin{equation}
  H_{(g)} = G_{(g)}^I \bar{p}_I
\end{equation}
is expressed in terms of secondary constraints only:
\begin{equation}
\label{eq:gconstraints}
  \{\bar{q}^I, \intd{\diff{x^1}} H_{(g)} \} = G_{(g)}^I = \partial q^I + P^{IJ} p_J 
\end{equation}

The extension to include conformal matter is straightforward. From the action
\eqref{eq:cmMFS} together with the matter fields
\begin{align}
  \matq &= f & \matq^\alpha &= \lambda^\alpha
\label{eq:lalapetz}
\end{align}
the canonical momenta\footnote{In what follows, all expressions are written in
terms of canonical variables. The determinant $\sqrt{-g}=e$ in these variables reads $e
= p_{--} \bar{p}_{++} - p_{++} \bar{p}_{--}$.}
\begin{align}
  \begin{split}
\label{eq:bosonP}
    \derfrac[L_{(m)}]{\dot{\matq}} = \matp&= \inv{e} \Bigl( ( p_{++} \bar{p}_{--} + p_{--} \bar{p}_{++} ) \partial \matq -
    2 p_{++} p_{--} \dot{\matq} \medsp
     &\quad + 2 i \bigl(p_{++} p_{--} (\bar{p}_- \matq^- + \bar{p}_+
    \matq^+) - \bar{p}_{++} p_{--} p_+ \matq^+ - p_{++} \bar{p}_{--}
    p_- \matq^- \bigr) \Bigr)\ ,
  \end{split}\medsp
\label{eq:P+}
\derfrac[L_{(m)}]{\dot{\matq}^+} =  \matp_+ &= - \inv{\sqrt{2}} p_{++} \matq^+\ ,\medsp
\label{eq:P-}
\derfrac[L_{(m)}]{\dot{\matq}^-} = \matp_- &= \inv{\sqrt{2}} p_{--}  \matq^-
\end{align}
are obtained. Analogous to \eqref{eq:hamdensity} the Hamiltonian density from
the matter Lagrangian in eq.\ \eqref{eq:cmMFS} is defined as
\begin{equation}
  \label{eq:matterham}
  H_{(m)} = \dot{\matq} \matp + \dot{\matq}^+ \matp_+ + \dot{\matq}^- \matp_- - L_{(m)}\ ,
\end{equation}
and the total Hamiltonian density is the sum of \eqref{eq:hamdensity} and
\eqref{eq:matterham}. Obviously the definition in \eqref{eq:matterham} implies
a Poisson bracket for the matter fields with the same structure as given in
\eqref{eq:canonicalbr}. Especially for field monomials one finds
\begin{align}
\label{eq:poissonbr2}
  \{ \matq, \matp' \} &= \delta(x - x') & \{ \matq^\alpha, \matp'_\beta\} &= -
  \delta^\alpha_\beta  \delta(x - x')\ .
\end{align}
We do not provide the explicit form of the matter
Hamiltonian, as it can again be written in terms of secondary constraints:
\begin{align}
\label{eq:classicalham}
  H &= G^I \bar{p}_I & G^I &= G^I_{(g)} + G^I_{(m)} & \{\bar{q}^I, \intd{\diff{x^1}} H_{(m)} \} = G_{(m)}^I
\end{align}
The explicit expressions for the matter part of the secondary constraints read:
\begin{align}
\label{eq:G++m}
  G^{++}_{(m)} &= - \inv{4 p_{++}} (\partial \matq - \matp)^2 + \inv{\sqrt{2}} \matq^+
  \partial \matq^+ + \frac{i}{p_{++}} (\partial \matq - \matp) p_+ \matq^+
  \medsp
\label{eq:G--m}
   G^{--}_{(m)} &= \inv{4 p_{--}} (\partial \matq + \matp)^2 - \inv{\sqrt{2}} \matq^-
  \partial \matq^- - \frac{i}{p_{--}} (\partial \matq + \matp) p_- \matq^-
  \medsp
\label{eq:G+m}
  G^+_{(m)} &= i (\partial \matq - \matp) \matq^+ \medsp
\label{eq:G-m}
  G^-_{(m)} &= -i (\partial \matq + \matp) \matq^-
\end{align}

\subsection{Second class constraints}

As the kinetic term of the matter fermion $\lambda$ is first order only, this
part of the action leads to constraints as well. From \eqref{eq:P+} and \eqref{eq:P-} the usual primary
second-class constraints are deduced:
\begin{align}
\label{eq:Psiconstr+}
   \Psi_+ &= \matp_+ + \inv{\sqrt{2}} p_{++} \matq^+ \approx 0\medsp
\label{eq:Psiconstr-}
   \Psi_- &= \matp_- - \inv{\sqrt{2}} p_{--} \matq^- \approx 0
\end{align}
These second class constraints are treated by substituting the Poisson bracket
by the ``Dirac bracket'' \cite{Dirac:1996,Gitman:1990,Henneaux:1992}
\begin{equation}
  \label{eq:diracbr}
  \{ f, g \}^* = \{ f, g \} - \{f, \Psi_\alpha\} C^{\alpha \beta} \{\Psi_\beta,
  g\}\ ,
\end{equation}
where
\begin{align}
  C^{\alpha \beta} C_{\beta \gamma} &= \delta^\alpha_\gamma\ , & C_{\alpha \beta}
  = \{\Psi_\alpha, \Psi_\beta\}\ .
\end{align}
From \eqref{eq:Psiconstr+} and \eqref{eq:Psiconstr-} together with the
definition of the canonical bracket the matrix $C_{\alpha \beta}$ follows as
\begin{equation}
  \label{eq:Cmatrix}
  C_{\alpha \beta} = \sqrt{2} \begin{pmatrix} - p_{++} & 0 \\ 0 & p_{--}
  \end{pmatrix}\ .
\end{equation}
In particular, the Dirac bracket among two field monomials has the non-trivial
components
\begin{align}
\label{eq:dirbr1}
  \{\matq^+, {\matq'}^+ \}^* &= \inv{\sqrt{2} p_{++}} \delta(x - x')\ , & \{\matq^-, \matq^- \}^* &=
  - \inv{\sqrt{2} p_{--}} \delta(x - x') \ , \medsp
\label{eq:dirbr2}
  \{\matp_+, \matp'_+\}^* &= \frac{p_{++}}{2 \sqrt{2}} \delta(x - x') \ , & \{\matp_-,
  \matp'_-\}^* &= -\frac{p_{--}}{2 \sqrt{2}} \delta(x - x') \ , \medsp
\label{eq:dirbr3}
  \{\matq^+,\matp'_+\}^* &= - \half{1} \delta(x - x') \ , & \{\matq^-,\matp'_-\}^* &= - \half{1}
  \delta(x - x') \ , \medsp
\label{eq:dirbr4}
  \{q^{++}, \matp'_+\}^* &= - \frac{\matq^+}{2\sqrt{2}} \delta(x - x') \ , & \{q^{--},
  \matp'_-\}^* &= \frac{\matq^-}{2\sqrt{2}} \delta(x - x') \ , \medsp
\label{eq:dirbr5}
  \{q^{++}, {\matq'}^+\}^* &= - \frac{\matq^+}{2 p_{++}} \delta(x - x') \ , & \{q^{--},
  {\matq'}^-\}^* &= - \frac{\matq^-}{2p_{--}} \delta(x - x')\ ,
\end{align}
and all other brackets are unchanged.

\subsection{Dirac bracket algebra of secondary constraints}

The algebra of Hamiltonian
constraints is the result of a straightforward but tedious calculation. We make a
few remarks on important observations therein.

As the $G^I_{(m)}$ are independent of $q^I$, the $p_I$ in
\eqref{eq:gconstraints} commutes trivially within $\{G^I_{(g)}, G^J_{(m)}
\}^*$. There are, however,  non-trivial commutators
with $q^I$: \eqref{eq:G++m} and \eqref{eq:G--m} depend on the coordinates $p_{\pm}$
and all constraints \eqref{eq:G++m}-\eqref{eq:G-m} depend on the fermionic matter
field $\matq^{\pm}$, which has a non-vanishing Dirac bracket with $q^{\pm\pm}$
\eqref{eq:dirbr5}. Indeed, after a straightforward calculation one arrives at
\begin{align}
\label{eq:constralgm1}
  \{ G^{++}_{(m)}, q'^{++} \}^* &= \inv{p_{++}} G^{++}_{(m)} + \inv{2 p_{++}^2} p_+
  G^+_{(m)} \delta(x - x')\ , \medsp
\label{eq:constralgm2}
  \{ G^{--}_{(m)}, q'^{--} \}^* &= \inv{p_{--}} G^{--}_{(m)} + \inv{2 p_{--}^2} p_-
  G^-_{(m)} \delta(x - x')\ ,
\end{align}
\begin{align}
\label{eq:constralgm3}
  \{ G^{++}_{(m)}, q'^{+} \}^* &= \inv{p_{++}} G^+_{(m)}\delta(x - x')\ , & \{ G^{--}_{(m)}, q'^{-} \}^* &=
  \inv{p_{--}} G^-_{(m)}\delta(x - x')\ , \medsp
\label{eq:constralgm4}
  \{ G^{+}_{(m)}, q'^{++} \}^* &= \inv{2 p_{++}} G^+_{(m)}\delta(x - x')\ , & \{ G^{-}_{(m)}, q'^{--} \}^* &= \inv{2 p_{--}} G^-_{(m)}\delta(x - x')\ ,
\end{align}
which are useful relations in the calculation of the ``mixed commutators'' $\{G^I_{(g)}, G^J_{(m)}
\}^*$. Also notice that $\inv{p_{++}} G^+_{(m)}$ is part of
$G^{++}_{(m)}$ as well.

However, the most important relations involved in the
calculation including the matter extension are the
supergravity restrictions \eqref{eq:sugraconstr1} and \eqref{eq:sugraconstr2}.
Also the fact that the ``supersymmetry transformation'' (the $P^{+|+}$,
resp.\ $P^{-|-}$ part of the Poisson tensor) is model independent leads to
cancellations between terms from the ``mixed'' commutators together with terms
from the purely matter part $\{G^I_{(m)}, G^J_{(m)} \}^*$.

Putting the pieces together the final result

\begin{align}
   \label{eq:constralg} \{ G^I, {G'}^J \}^* &=  G^K {C_K}^{IJ} \delta(x
   - x')\ , &
   {C_K}^{IJ} &= - \partial_K P^{IJ}
\end{align}

\noindent is obtained.
It is an
important confirmation of the construction of the last 
section that the result as expected from bosonic gravity is reproduced: Minimally coupled conformal
matter does not change the structure functions of the constraint
algebra as compared to the matterless case. Indeed, when Dirac brackets are
used, the result of \cite{Bergamin:2002ju} is found up to the
substitution $G_{(g)}^I \rightarrow G^I = G_{(g)}^I + G_{(m)}^I$.

Eq.\ (\ref{eq:constralg}) is the main result of this section and we would like to conclude with some comments on its algebraic structure. Beside the constraints $G^I$, the r.h.s.\ of this relation depends on the
canonical coordinates $q^I$. To study the closure of this non-linear algebra,
at least that coordinate must be part of the algebra as well. In
super-geometry it turns out that the algebra of $G_{(g)}^I$ and
$q^I$ closes as \cite{Bergamin:2002ju}
\begin{equation}
  \label{eq:constralg2}
  \{ G^I_{(g)}, {q'}^J \} = - P^{IJ} \delta(x
   - x')\ .
\end{equation}
Therefore the system $(G^I_{(g)}, q^I)$ defines a graded finite W algebra.
It simplifies to a graded Lie-algebra if and only if the Poisson tensor is
linear in the target-space coordinates. As long as only super-geometry
is considered, the relations \eqref{eq:constralg} and
\eqref{eq:constralg2} hold for any Poisson tensor.

In the presence of matter interaction \eqref{eq:cmMFS}, eq.\ \eqref{eq:constralg2}
must be replaced by $\{ G^I, q'^J \}^*$, and the complete algebra is
obtained by adding \eqref{eq:constralg} and
\eqref{eq:constralgm1}-\eqref{eq:constralgm4}. Notice that on the r.h.s.\ of
$\{ G^I, q'^J \}^*$ the terms $G^I_{(g)}$ and $G^I_{(m)}$ never appear together, because $P^{++|++} =
P^{++|+}=0$. As the r.h.s.\ of \eqref{eq:constralgm1}-\eqref{eq:constralgm4}
depend on $p_J$ as well, the algebra no longer closes together with $q^I$, but
the former coordinate must be part of the algebra as well. But
since \cite{Bergamin:2002ju}
\begin{equation}
  \{ G^I_{(g)}, p'_J \} = (-1)^I \partial \delta(x - x') \delta^I_J + (-1)^{IJ}
  \partial_J P^{IK} p_K \delta(x - x')\ ,
\end{equation}
this is not a W algebra, as expected from the analogous situation in bosonic case.

\section{Quantization}
\label{sec:3}
\subsection{Ghosts and gauge fixing}
The construction outlined in this section closely follows
ref.\ \cite{Henneaux:1992}. However, some details of the grading are different, as
the canonical index position in that book is different from the conventions used
here. Our constraints have upper indices and thus the same applies to the
anti-ghosts
to be introduced in relation to the constraints of our system:
\begin{align}
  \text{primary constraints:}&\ \ \ \ b_I \ \ p_b^I \medsp
  \text{secondary constraints:}&\ \ \ \ c_I \ \ p_c^I
\end{align}
The brackets between ghosts and anti-ghosts are defined conveniently as 
\begin{align}
  [ b_I, p_b^J ] &= - (-1)^{(I+1)(J+1)} [ p_b^J, b_I ] =  \delta_I^J\ , & [ c_I,
  p_c^J ] &= - (-1)^{(I+1)(J+1)} [ p_c^J, c_I ] =  \delta_I^J\ .
\end{align}
To first order in homological perturbation theory the BRST charge $\Om$ follows straightforwardly:
\begin{equation}
\label{eq:omegadef}
  \Omega = \bar{q}^I b_I + G^I c_I - \half{1} (-1)^I p_c^K \partial_K P^{IJ} c_J c_I
\end{equation}
Note that $P^{IJ} c_J c_I = (-1)^{I+J} P^{JI} c_I c_J$.

For the special case $G^I = G^I_{(g)}$ the BRST charge \eqref{eq:omegadef} is found to be
nilpotent, simply as a consequence of the graded Jacobi identity\footnote{Without
  coupling of matter fields $\Omega^2 = 0$ does not depend on the specific
  form of the Poisson tensor in \eqref{eq:mostgensup}-\eqref{eq:mostgensuplast}.} of $P^{IJ}$ \eqref{eq:nijenhuis}. But even
for the constraints involving  $G^I_{(m)}$ the homological perturbation theory stops at
this order: For models with $Z = 0$ in \eqref{eq:mostgenaction} all relevant
brackets $[ G^I, \partial_L P^{JK} ]^*$ vanish trivially, for the general case
the restrictions \eqref{eq:sugraconstr2} guarantee nilpotency.

As the Hamiltonian vanishes on the constraint surface it simply becomes
\begin{equation}
  \label{eq:quant10}
  \gf{H} = \{ \Omega, \Psi \}
\end{equation}
for some gauge fixing fermion $\Psi$. Following \cite{Grumiller:2001ea} a multiplier
gauge $\Psi =  p_c^I a_I$ is used with constant $a_I$, for which the gauge-fixed Hamiltonian reads 
\begin{equation}
\label{eq:quant11}
  \{ \Omega, \Psi \} = - G^I a_I + (-1)^K p_c^I \partial_I P^{JK}
  c_K a_J\ .
\end{equation}
The canonical equations
of the ghosts are
\begin{align}
  \varfrac[\gf{H}]{p_c^I} &= \dot{c}_I\ , & \varfrac[\gf{H}]{c_I} &= (-1)^I \dot{p}_c^I\ ,
\end{align}
and the corresponding gauge-fixed Lagrangian reads
\begin{equation}
\label{eq:quant11.2}
  \gf{L} = \dot{\bar{q}}^I\bar{p}_I + \dot{q}^I p_I + \dot{\matq} \matp + \dot{\matq}^\alpha \matp_\alpha
  + p_c^I \dot{c}_I + p_b^I \dot{b}_I - \gf{H}\ .
\end{equation}

Even in such a simple gauge the construction of the effective
Lagrangian and the subsequent integration over its variables in the
path integral can lead to
lengthy equations. Thus we restrict the explicit calculations to the simplest
possible gauge, which is not inconsistent with physical requirements such as
the non-degeneracy of the bosonic metric\footnote{Notice that
  according to the notation \eqref{eq:A10} $p_{++}$ is purely imaginary. Also, one
  component ($q^+$, $\matq^-$) of a spinor is real, while the other ($q^-$, $\matq^+$) is imaginary.}: $a_I =
- i {\delta_I}^{++}$. This entails the Eddington-Finkelstein form of the bosonic line element.

The gauge-fixed Lagrangian \eqref{eq:quant10}-\eqref{eq:quant11.2} with this
choice becomes
\begin{equation}
  \label{eq:quant12}
\begin{split}
  \gf{L} &= \dot{\bar{q}}^I\bar{p}_I + \dot{q}^I p_I + \dot{\matq} \matp + \dot{\matq}^\alpha
  \matp_\alpha + p_c^I \dot{c}_I + p_b^I \dot{b}_I - i \partial q^{++}
  - i P^{++|J} p_J \medsp
  &\quad + \frac{i}{4 p_{++}} (\partial \matq - \matp)^2 - \frac{i}{\sqrt{2}} \matq^+
  \partial \matq^{+} + \frac{1}{p_{++}}(\partial \matq - \matp) p_+ \matq^+ \medsp
  &\quad + i (-1)^K p_c^I \partial_I P^{++|K} c_K\ ,
\end{split}
\end{equation}
which will be the starting point of our calculations in the path
integral formalism.

\subsection{Path integral formalism}
Before formally integrating out some of the fields by the path integral
formalism it is worthwhile to check that no problems arise due to ordering
ambiguities hidden in the definition of its measure (which are typical for
nonlinear interactions, cf.\ e.g.\ \cite{Bastianelli:1993ct}) and that
boundary contributions are treated with care. Since these considerations are
somewhat tangential to the main topics of this paper, they are relegated to
appendix \ref{app:B}, where it is shown that problems of this type do not arise.

The generating functional of Green functions that follows from the result of the previous section
has to be integrated over all physical fields and all ghosts. Together with
sources $\jq_I$, $\jp^I$ for the geometrical variables and $J$, $J_\alpha$ for
the matter fields it reads:
\begin{equation}
  \label{eq:quant13}
  \begin{split}
    \mathcal{W}[\jq_I, \jp^I,J,J_\alpha] &= \int
    \mathcal{D}(q^I,p_I,\bar{q}^I,\bar{p}_I,\matq,\matp,\matq^\alpha,\matp_{\alpha},c_I,p_c^I,b_I,p_b^I)\medsp
    &\quad \cdot \exp \Bigl( i \intd{\diff{^2 x}} \bigl(\gf{L} + q^I \jq_I + \jp^I
    p_I + \matq J + \matq^{\alpha} J_\alpha \bigr)\Bigr)
  \end{split}
\end{equation}
The gauge-fixed Hamiltonian, being independent of $\bar{q}^I$, $\bar{p}_I$, $b_I$ and
$p_b^I$, allows a trivial integration of all these fields. As the remaining
ghosts appear at most bi-linearly in the action they can be integrated
over, which leads to the super-determinant
\begin{equation}
  \label{eq:quant14}
  \mbox{sdet}  {M_I}^J = \mbox{sdet} \bigl( {\delta_I}^J \partial_0 -
  i \partial_I P^{++|J} \bigr)\ .
\end{equation}
The integration of $\matp_\alpha$ by means of the constraint \eqref{eq:Psiconstr+}
and \eqref{eq:Psiconstr-} is trivial as well, while the bosonic momentum
$\matp$ of matter
can be integrated after the quadratic completion
\begin{equation}
  \label{eq:quant15}
\begin{split}
  L_{(m)} &= \dot{\matq} \matp + \dot{\matq}^\alpha \matp_\alpha + \frac{i}{4
  p_{++}} (\partial \matq - \matp)^2 - \frac{i}{\sqrt{2}} \matq^+
  \partial \matq^{+} + \frac{1}{p_{++}}(\partial \matq - \matp) p_+ \matq^+ \medsp
  &= \frac{i}{4 p_{++}} \bigl( \matp - \partial \matq + 2 i p_+ \matq^+ - 2
  i p_{++}\dot{\matq} \bigr)^2 + i
  p_{++} \dot{\matq}^2 + \dot{\matq}(\partial \matq - 2 i p_+ \matq^{+}) 
  \medsp
  &\quad - \inv{\sqrt{2}} p_{++} \dot{\matq}^+ \matq^{+} + \inv{\sqrt{2}} p_{--}
  \dot{\matq}^- \matq^{-} - \frac{i}{\sqrt{2}} \matq^+ \partial \matq^+
\end{split}
\end{equation}
and a suitable shift in this variable. After a Gaussian integration in the
shifted $\matp$ this leads to a new effective Lagrangian linear in
$p_I$, which allows to integrate these fields as well. The resulting functional
$\delta$-functions finally determine the $q^I$. It is advantageous to proceed through
these steps separately for the matterless and for the full theory.
\subsection{Matterless case}

\subsubsection{Integrating out geometry}
After integrating the remaining ghosts the Lagrangian
\begin{equation}
  \label{eq:quant16}
  \itindex{L}{eff} = \dot{q}^I p_I - i \partial q^{++} - i P^{++|J} p_J + q^I \jq_I + \jp^I
    p_I
\end{equation}
is obtained. The $p_I$ only appear
\emph{linearly} in \eqref{eq:quant16} which upon integration yields five
functional $\delta$ functions. The arguments of the latter imply the solution
of a system of as many coupled differential equations, which for the Poisson
tensor \eqref{eq:lorentzcov},
\eqref{eq:mostgensup}-\eqref{eq:mostgensuplast} become ($W = - e^Q
u^2/8$, cf.\ eq.\ \eqref{eq:bosonicC})
\begin{align}
  \dot{q}^\phi &= - i q^{++} - \jp^\phi\ , \label{eq:quant17.1}\medsp
  \dot{q}^{++} &= - \jp^{++}\ , \label{eq:quant17.2}\medsp
  \dot{q}^{--} &= i \Bigl( e^{-Q} W' + q^{++} q^{--} Z + \inv{2 \sqrt{2}} q^- q^+
  e^{-Q/2}\bigl( \sqrt{-W} \bigr)''\Bigr)- \jp^{--}\ ,
  \label{eq:quant17.3}\medsp
  \dot{q}^{+} &= - \jp^{+}\ , \label{eq:quant17.4}\medsp
  \dot{q}^{-} &= - i \bigl(e^{-Q/2}(\sqrt{-W})' q^+ - \half{1} Z q^{++}
  q^- \bigr) - \jp^{-}\ , \label{eq:quant17.5}
\end{align}
and can be used to perform the final $q^I$
integration. As expected, this exactly cancels the super-determinant \eqref{eq:quant14}.
The solutions of \eqref{eq:quant17.1}-\eqref{eq:quant17.5} are denoted by
$B^I$ and may formally be written as
\begin{equation}
  B^\phi = \hat{q}^\phi + \partial_0^{-1} A^\phi \ ,
  \label{eq:quant18.1}
\end{equation}\vspace{-3ex}
\begin{align}
\label{eq:quant18.2}
  B^{++} &= \hat{q}^{++} + \partial_0^{-1} A^{++}\ , & 
  B^{--} &= e^{-D^{--}}\bigl(\hat{q}^{--}  + \partial_0^{-1} A^{--} \bigr)\ , \medsp
\label{eq:quant18.5}
 B^{+} &= \hat{q}^{+} + \partial_0^{-1} A^+ \ , & B^{-} &=
 e^{-D^{--}/2}\bigl(\hat{q}^{-} +  \partial_0^{-1} A^- \bigr)\ ,
\end{align}
with
\begin{align}
A^\phi &= -iB^{++} - \jp^\phi\,, \label{eq:quant1783.1} \\
A^{++} &= -\jp^{++}\,, \label{eq:quant1783.2} \\
A^{--} &= e^{D^{--}}\left[ie^{-Q}W'+\frac{i}{2\sqrt{2}}e^{-Q/2}(\sqrt{-W})''B^-B^+ - \jp^{--}\right]\,, \label{eq:quant1783.3}\\
A^+ &= - \jp^+\,, \label{eq:quant1783.4}\\
A^- &= -e^{D^{--}/2}\left[ie^{-Q/2}(\sqrt{-W})'B^+ + \jp^-\right]\,.\label{eq:quant1783.5}
\end{align}
Here the $\hat{q}^I(x^1)$ are integration constants $\partial_0 \hat{q}^I = 0$,
$\partial_0^{-1}$ denotes a properly defined Green function and $D^{--} =
i \partial_0^{-1}(Z B^{++})$.
With these definitions the solution for the matterless path integral finally
becomes
\begin{equation}
  \label{eq:quant19}
  \mathcal{W}[\jp^I, \jq_I] = \exp{iL^0_{\rm eff}}\,, \quad L^0_{\rm eff}=\intd{\diff{^2x}} \left(B^I \jq_I + \tilde{L}^0 (\jp^I, B^I)\right) \ .
\end{equation}
$\tilde{L}^0$ are the so-called ``ambiguous terms''. An expression of this
type is generated by the integration constants $G_I(x^1)$ from the term $\intd{\diff
  x^0}\intd{\diff y^0} (\partial_0^{-1}A^I) \jq_I$ (cf.\ section 7 of \cite{Grumiller:2002nm} and references
therein). Thus, the ambiguous terms are
\begin{equation}
  \label{eq:quant666}
  \tilde{L}^0 = A^Ig_I\,.
\end{equation}
with $A^I$ given by \eqref{eq:quant1783.1}-\eqref{eq:quant1783.5}.
The physical meaning of the integrations constants $g_I$ can be seen from the
expectation values $<p_I>$ which have to be adjusted in accordance with the
boundary conditions imposed on these fields. For instance, the expectations
values $<\psi_1^\pm>$ are required to vanish asymptotically and hence the
corresponding constants $g_\pm$ have to be set to zero. Then the matter
vertices to be calculated below will be generated solely by $A^{--}$, just as
in the bosonic case.

\subsubsection{Effective action and local quantum triviality}

It is instructive to calculate the Legendre transform of $L^0_{\rm eff}$, viz.\ the effective action
\begin{equation}
  \label{eq:745}
  \Ga\left(<q^I>,<p_I>\right):=L^0_{\rm eff}\left(\jq_I,\jp^I\right)-\intd{\diff{^2x}}\left(<q^I>\jq_I+\jp^I<p_I>\right)\,,
\end{equation}
in terms of the mean fields
\begin{equation}
  \label{eq:668}
  <p_I>:=\left. \frac{\stackrel{\rightarrow}{\delta}}{\de \jp^I} L^0_{\rm
  eff}\right|_{j=0}\ ,\quad<q^I>:=\left. L^0_{\rm eff} \frac{\stackrel{\leftarrow}{\delta}}{\de \jq_I} \right|_{j=0} = \left. B^I\right|_{\jp=0}\,.
\end{equation}
To ensure that the mean fields attain their classical values without supersymmetry signs, left variation is used in the left formula of (\ref{eq:668}) and right variation in the right one.
Plugging in the previous results \eqref{eq:quant19}, \eqref{eq:quant666} of this section immediately yields
\begin{equation}
  \label{eq:669}
  \Ga\left(<q^I>, <p_I>\right)= \intd{\diff{^2x}} \left(A^Ig_I-\jp^I<p_I>\right)\,,
\end{equation}
where the sources $\jp^I$ have to be expressed in terms of the (classical)
target space coordinates and their first derivatives by virtue of
(\ref{eq:quant17.1})-(\ref{eq:quant17.5}). Consequently, the effective action
(\ref{eq:669}) turns out to be nothing but the gauge fixed {\em classical}
action \eqref{eq:quant16} in terms of the mean fields plus the nontrivial boundary
contributions\footnote{The boundary terms result from $A^Ig_I$, so actually
  there are five of them. However, three of them are rather trivial, as can be
  checked most easily by expressing the sources in
  (\ref{eq:quant1783.1})-(\ref{eq:quant1783.5}) in terms of the classical
  target space coordinates using again
  (\ref{eq:quant17.1})-(\ref{eq:quant17.5}).}
\begin{equation}
  \label{eq:670}
  \pm \int_{\partial M}\!\!\!\!\!\!\diff{x^1} \left(e^{D^{--}}B^{--}g_{--}-e^{D^{--}/2}B^-g_-\right)\,.
\end{equation}
Requiring asymptotically vanishing fermion fields (cf.\  section \ref{sec:vertices.1} below
eq.\ \eqref{eq:quant24}), $g_-=0$ and the second term vanishes. The sign in
front of \eqref{eq:670} depends on the sign of the outward pointing normal
vector of the boundary surface $\partial M$. Thus, local quantum triviality
generalizes from the bosonic case \cite{Kummer:1997hy} to the supersymmetric
one.

\subsubsection{Nonlocal quantum correlators}
Before switching on matter interactions we would like to address a further
point, which already has been observed (but not discussed) in the bosonic case
\cite{Grumiller:2003dh}: although all canonical variables acquire their
classical values as their expectation values, the correlators between two or
more variables need not decompose into a product of classical values---the
latter feature being very well-known from generic quantum field theory. In particular, we obtain
\begin{equation}
  \label{eq:671}
  <q^Iq^J>=<q^I><q^J>\ ,\quad<p_Ip_J>\neq<p_I><p_J>\ ,
\end{equation}
and according to \eqref{eq:quant18.1}-\eqref{eq:quant18.5}
\begin{equation}
  \label{eq:671.1}
  <q^I(x)p_J(y)>=<q^I(x)><p_J(y)>+\frac{\de }{i\de \jp^J(y)} B^I(x)\ .
\end{equation}
Eq.\ \eqref{eq:671} e.g.\ entails that the product of two zweibein
components is not necessarily equal to the product of the vacuum expectation
values thereof. Note that in order to create nontrivial correlators of this
type the dependence on external sources must be non-linear. This rules out
many of the possible correlators. Among the expressions \eqref{eq:671.1} with
purely bosonic field content
\begin{equation}
  \label{eq:671.2}
  <X^a(x) e_{a 1}(y)> = < q^{++}(x) p_{++}(y) + q^{--}(x) p_{--}(y) >
\end{equation}
is the only Lorentz invariant contribution that receives non-perturbative quantum
corrections. In the coincidence limit $x=y$ \eqref{eq:671.2} reduces to a
1-form times the Killing norm.

Even simpler are correlators of fermionic operators. According to our boundary
conditions employed below (cf.\ sect.\ \ref{sec:vertices.1})
\begin{equation}
  \label{eq:672}
  <\chi^\pm>=0=<\psi_\pm>
\end{equation}
occurs. This implies that all correlators involving one or more insertions of
a dilatino or a gravitino vanish, with the notable exception of
$<\chi^+(x)\psi_{+1}(y)>$. This one turns out to be especially useful to study the
systematics of the non-perturbative quantum effects: In contrast to
\eqref{eq:671.2} the classical (``dominant'') value vanishes from
\eqref{eq:672} and consequently any nontrivial contribution stems exclusively from
quantum effects.

To calculate this correlator explicitly we should impose certain
prescriptions for the integration constants in \eqref{eq:671.1}. With
$\partial_0^{-1}{}_{xz}f(z):=\int^x_yf(z)\diff z$ and $\theta(0)=1/2$ it becomes\footnote{\label{foot123}Other prescriptions are possible, of course, but eventual ambiguities in the definitions are fixed by the (physical) requirement that in the coincidence limit the correlator vanishes, because after all the theory is {\em locally} quantum trivial. The choice $\theta(0)=1/2$ has the advantage that the identity $\theta(a-b)=1-\theta(b-a)$ can be used even if $a=b$.  The results of this
  section on nonlocal correlators ($a \neq b$) do not depend on the choice of
  the prescription.} the field independent expression
\begin{equation}
  \label{eq:673}
  <\chi^+(x)\psi_{+1}(y)>=\frac{\de}{i\de \jp^+(y)}
  B^+(x)=\frac1i\left(\theta(y^0-x^0)-\frac12\right)\de(x^1-y^1)\ ,
\end{equation}
where the lower boundary $y$ is chosen such that in the coincidence limit
$x^0=y^0$ of (\ref{eq:673}) the correlator vanishes. This correlator is purely nonlocal in
$(x^0,y^0)$, but local in $(x^1,y^1)$ and obeys the
following identities (we recall that $\psi_{\pm 0}=0$):
\begin{equation}
\label{eq:identities}
<\chi^+(x)\psi_{+m}(y)>=<\chi^-(x)\psi_{-m}(y)>\,,\quad<\chi^\pm(y)\psi_{\pm
  m}(x)>=<\chi^\pm(x)\psi_{\pm m}(y)>\dega \ .
\end{equation}
Therefore, the matrix
\begin{equation}
  \label{eq:674}
  \left(\begin{array}{cc}
   <\chi^+(x)\chi_+(y)> & <\psi^+_1(x)\chi_+(y)> \\
   <\chi^+(x)\psi_{+1}(y)> & <\psi^+_1(x)\psi_{+1}(y)>    
  \end{array}\right)
\end{equation}
has non-vanishing entries in the off-diagonal only. There is a mixed gravitino-dilatino correlator which is completely independent of the geometric properties encoded in the bosonic potentials $u$ and $Z$.
While the pseudo-scalar expression
\begin{equation}
<\chi^\al(x)\gthree\psi_{\al m}(y)>=<\chi^+(x)\psi_{+m}(y)>-<\chi^-(x)\psi_{-m}(y)>=0
\label{eq:chiralcondensate}
\end{equation}
vanishes, the scalar one is non-vanishing ($\eps(a-b):=\theta(a-b)-\theta(b-a)$):
\begin{equation}
<\chi^\al(x)\psi_{\al m}(y)>=\frac1i \eps(y^0-x^0)\de(x^1-y^1)\de_{1m}
\label{eq:straightcondensate}
\end{equation}

In order to get a gauge independent\footnote{The result of our construction
  resembles a Wilson loop. While we have some reservations regarding the
  observability of a generic Wilson loop (as opposed to certain quantities
  derived from it like the anomalous dimension), cf.\ e.g.\
  \cite{Grumiller:2003sk}, in the present context the attribute ``gauge
  independent'' is justified because the correlator does not receive any
  contributions from renormalization as our derivation has been an exact one
  rather than perturbative.} expression one has to consider the correlator integrated over a path $\mathcal{P}$,
\FIGURE[t]{\parbox{\linewidth}{\centering\epsfig{file=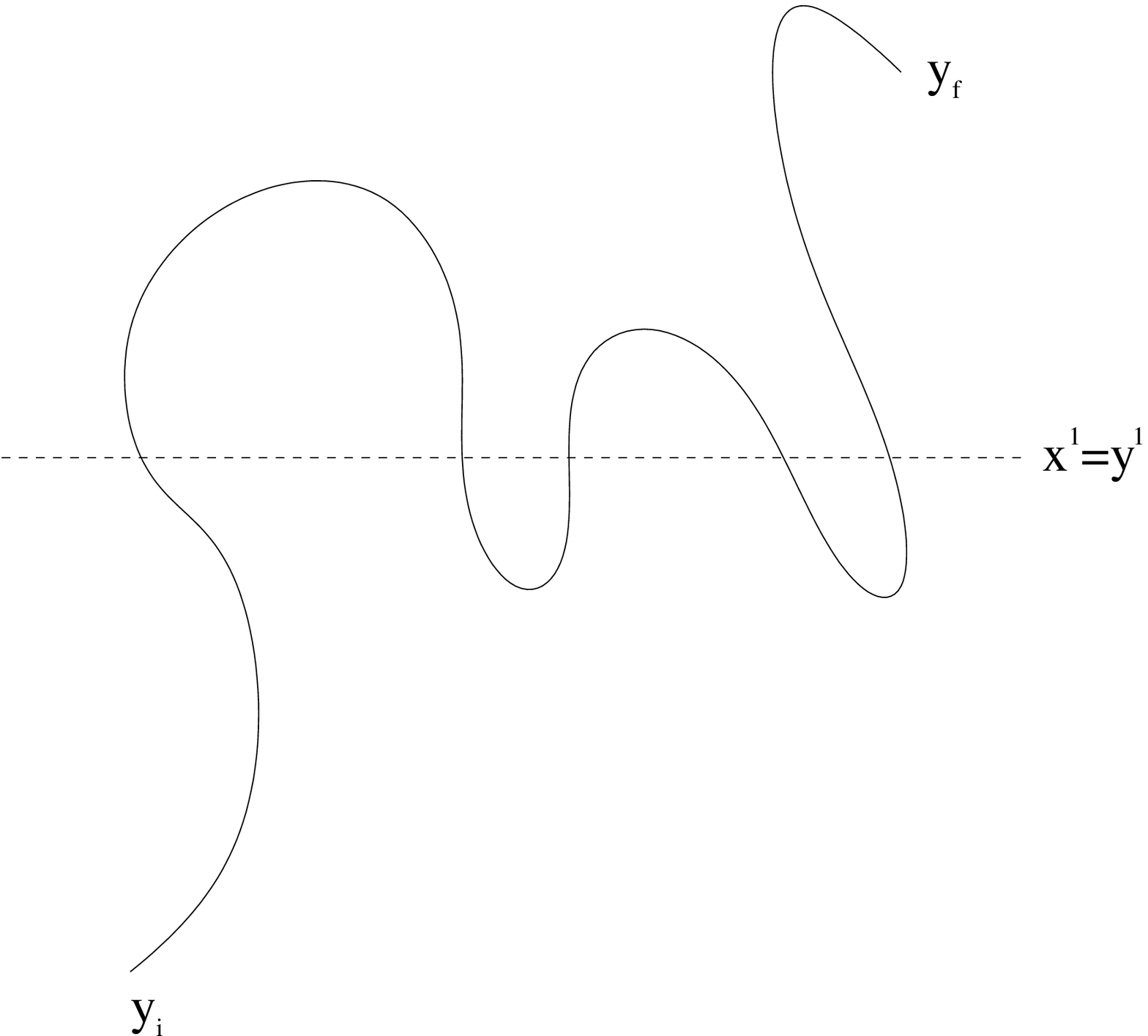,width=0.4\linewidth}}
\caption{A typical path $\mathcal{P}$ with $n_+^r=3$ and $n_-^r=2$. It is supposed that the point $x$ is on the left side of this graph, i.e.\ $x^0<y^0$ for all $y$ lying on $\mathcal{P}$. Thus, $\mathcal{K}=3-2-0+0=1$}
\label{fig:1}}
\begin{equation}
\mathcal{K}:=i\Big<\int_{\mathcal{P}}\!\!\!\diff y^m \chi^\al(x)\psi_{\al m}(y)\Big>=\int_{\mathcal{P}}\!\!\!\diff y^1\eps(y^0-x^0)\de(x^1-y^1)=(n_+^r - n_-^r- n_+^l + n_-^l)\,.
\label{eq:intcondensate}
\end{equation}
The natural numbers $n_\pm^{rl}$ count how often the line $y^1=x^1$ is
intersected from below ($+$) and from above ($-$) by the path $\mathcal{P}$. The
additional label $r,l$ contains the information whether the intersection point
lies on the left or the right hand side of $x$ (``left'' means
$x^0>y^0$). This can be visualized most easily by plotting the path in a
Cartesian diagram with x-axis $(y^0-x^0)$ and y-axis $(y^1-x^1)$. Any
intersection of the path $\mathcal{P}$ with the y-axis from an even (odd)
quadrant to an odd (even) one contributes $+1$ ($-1$) to
$\mathcal{K}$. Intersections through the origin do not contribute. The
integrated correlator $\mathcal{K}$ takes values in $\mathbb{Z}$. Obviously,
$\mathcal{K}$ is independent of the choice of integration constants as long as
the Wilson loop does not pass through the origin (cf.\ footnote \ref{foot123}).

\FIGURE[t]{\parbox{\linewidth}{
\centering
\epsfig{file=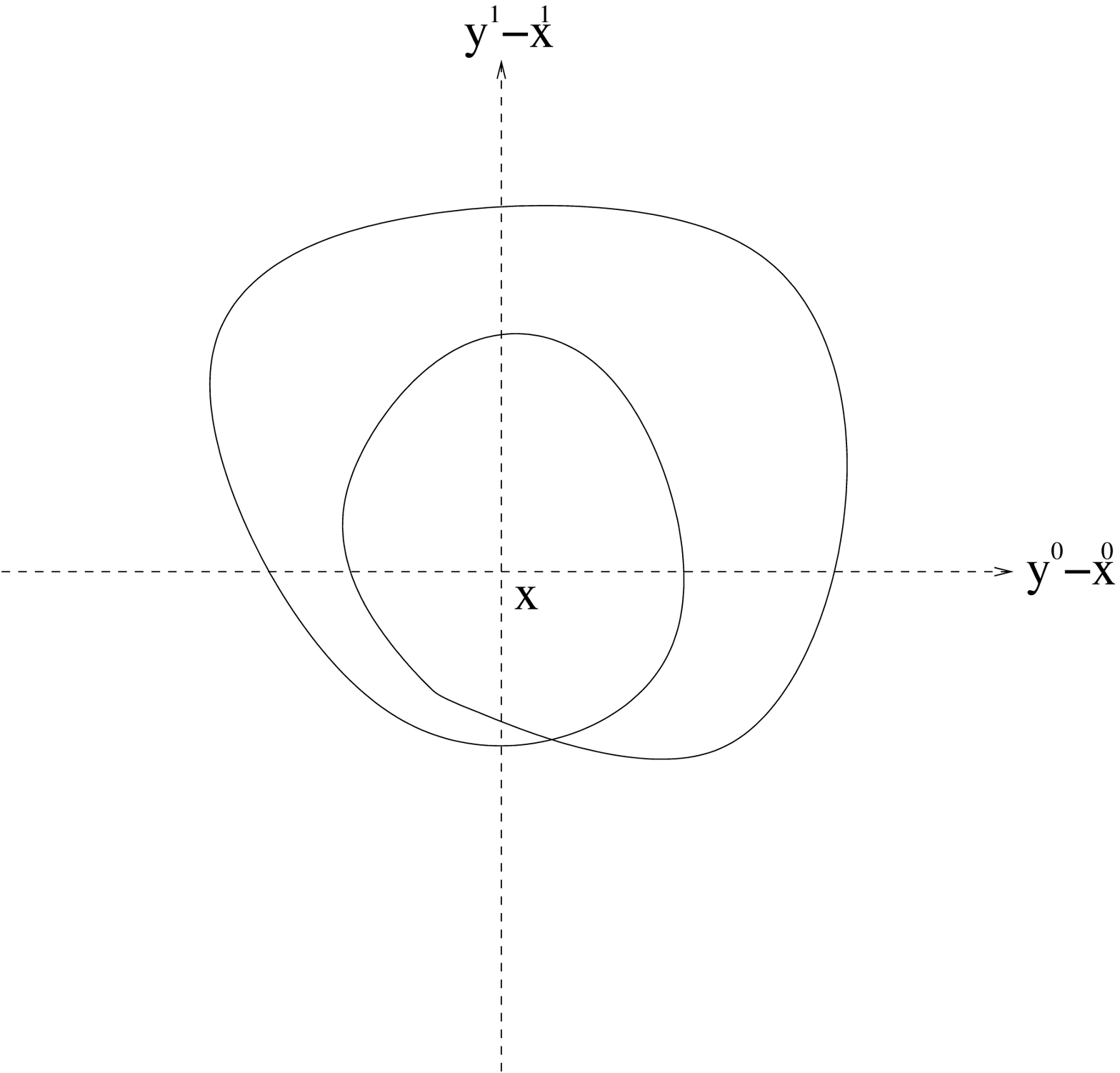,width=0.4\linewidth}
\caption{A closed path winding around the point $x$ with $\mathcal{K}=\pm 4$. The sign depends on the orientation and it is positive for counter clockwise orientation.}
\label{fig:2}}}
For strip-like topology a typical open path from the point $y_i$ to $y_f$ is
displayed in fig.\ \ref{fig:1}. An example for a closed path with nontrivial
winding is given in fig.\ \ref{fig:2}. Obviously, $\mathcal{K}/2$ can be
interpreted as winding number around the point $x$. For cylindrical topology
(fig.\ \ref{fig:3}) further complications are possible: if $\mathcal{K}/2$ is
again interpreted as winding number this implies that cylindrical topology
allows for a fractional (half-integer) winding number. The upper and lower boundaries have to be identified in that graph. The non-compact direction corresponds to ``time'', the compact one to the ``radius'' $x^0$. Note that $x^1$ is light-like (typically the retarded time).
\FIGURE[t]{\parbox{\linewidth}{
\centering
\epsfig{file=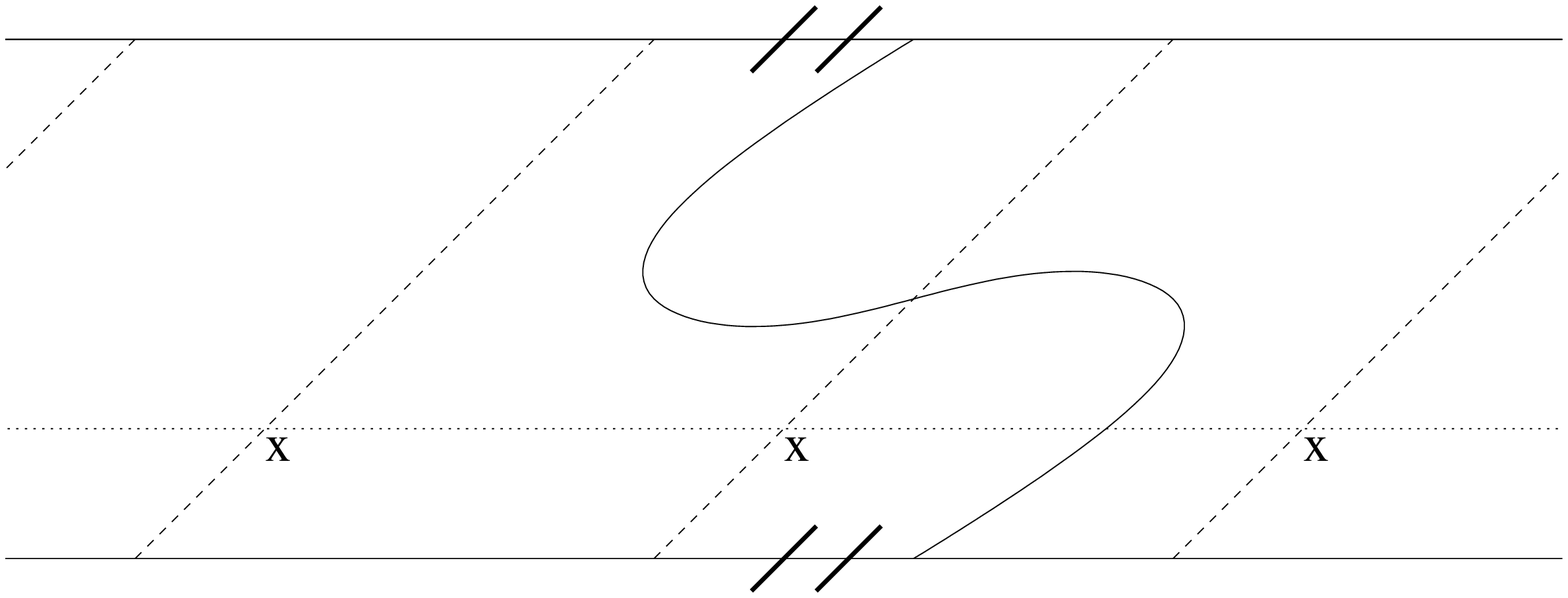,width=0.5\linewidth}
\caption{A nontrivial closed path for cylindrical topology with $\mathcal{K}=\pm 1$, depending again on the orientation of the loop.}
\label{fig:3}}}

In conclusion, the quantity $\mathcal{K}\in\mathbb{Z}$ is a topological
invariant, because it is completely independent of the metric and of the
conformal frame. $\mathcal{K}/2$ can be interpreted as a winding number of a
given path $\mathcal{P}$ around a reference point $x$. It is emphasized that
this is {\em not} the case for all other nontrivial correlators of type
(\ref{eq:671.1}).  Indeed, $< X^{a}(x) e_{a}(y) >$ in eq.\ \eqref{eq:671.2} receives similar
contributions, but as its classical contribution from eq.\
\eqref{eq:671.1} does not vanish the corresponding Wilson line is not a
topological invariant, unless $x$ coincides with a bifurcation point
($X^a = 0$).

The discussion above is in accordance with local quantum triviality but
possible nonlocal quantum non-triviality after exact (background independent)
quantum integration of the geometry.
\subsection{Matter interactions}
Although the result of the matterless case is an important consistency check of
the calculations, the main motivation of this approach to quantum gravity is
the study of matter interactions. Here the path integral after
performing the momentum integrations of the matter fields takes the form
\begin{equation}
  \label{eq:quant20}
  \mathcal{W}[\jq_I, \jp^I,J,J_\alpha] = \int \mathcal{D}(q^I,p_I,\matq,\matq^\alpha) (\det
  p_{++})^{1/2} \mbox{sdet} {M_I}^J \exp\Bigl(i \intd{\diff{^2 x}}
  (L_{(g)} + L_{(m)})\Bigr)\ .
\end{equation}
The determinant of $p_{++}$ originates from the shifted $\matp$ integration,
$L_{(g)}$ is the Lagrangian \eqref{eq:quant16} and
$L_{(m)}$ the $\matp$ independent part of \eqref{eq:quant15}
supplemented by the source terms in $J$ and $J^\alpha$.

We denote the right hand side of the differential equations
\eqref{eq:quant17.1}-\eqref{eq:quant17.5} by $\itindex{q}{(g)}^I$. Then the
related set of differential equations in presence of matter coupling can be
abbreviated as
\begin{align}
  \dot{q}^\phi &= q_{(g)}^{\phi}\ , \label{eq:quant21.1}\medsp
  \dot{q}^{++} &= q_{(g)}^{++} - i \dot{\matq}^2 + \inv{\sqrt{2}}\dot{\matq}^+
  \matq^+\ , \label{eq:quant21.2}\medsp
  \dot{q}^{--} &= q_{(g)}^{--} - \inv{\sqrt{2}} \dot{\matq}^-
  \matq^-\ ,
  \label{eq:quant21.3}\medsp
  \dot{q}^{+} &=  q_{(g)}^{+} - 2i \dot{\matq} \matq^+\ , \label{eq:quant21.4}\medsp
  \dot{q}^{-} &=  q_{(g)}^{-}\ . \label{eq:quant21.5}
\end{align}
The solutions of \eqref{eq:quant21.1}-\eqref{eq:quant21.5} are denoted by
$\hat{B}^I$ and the remaining path integral becomes
\begin{multline}
  \label{eq:quant22}
  \mathcal{W}[\jq_I, \jp^I,J,J_\alpha] = \int \mathcal{D}(\tilde{\matq},\tilde{\matq}^\alpha) \medsp
  \exp\Bigl(i \intd{\diff{^2x}} \dot{\matq} \partial \matq - \frac{i}{\sqrt{2}} \matq^+
  \partial \matq^+ + \hat{B}^I \jq_I + \matq J + \matq^\alpha J_\alpha +
  \tilde{L} (\jp^I, \hat{B}^I)\Bigr)\ ,
\end{multline}
with $\mathcal{D}(\tilde{\matq},\tilde{\matq}^\alpha)$ being the properly defined path
integral measure with convenient super-conformal properties
\cite{Toms:1987sh,Rocek:1986iz}. Its derivation is outlined in section
\ref{sec:loops}. The term $\tilde{L}$ is produced in the same way as
$\tilde{L}^0$ in \eqref{eq:quant19},\eqref{eq:quant666} for the matterless
case. Because for all physical Green functions the sources of the target space
coordinates are zero ($\jq_I = 0$) it actually  encodes all nontrivial interactions, and
in general is non-polynomial and nonlocal in the matter fields. Therefore, the
remaining integrations over them cannot be performed exactly and one has to
rely on a
perturbation expansion.
\section{Lowest order tree graphs}\label{sec:vertices}
\subsection{Localized matter}\label{sec:vertices.1}

To simplify the derivation of the lowest order vertices\footnote{Alternatively
  one can derive them by brute force methods encountering the difficulty how
  to define properly operators of the type $\partial_0^{-1}$. Although this
  can be done \cite{Kummer:1998zs} the calculations become rather lengthy already in the bosonic case. Therefore, only the simpler method described in the text will be presented, but it is emphasized that of course both methods give the same answer. For higher order $(2n+2)$-point vertices matter will have to be localized at $n$ different points, but here only the case $n=1$ will be studied.} the concept of
``localized matter'' is useful
\cite{Kummer:1998zs,Grumiller:2000ah,Fischer:2001vz}. To this end we introduce
for the localized matter at some point $y$ the notation
\begin{align}
  &\Phi_0(x) = \half{1} \dot{\matq}^2(x) \Rightarrow a_0 \delta^2(x-y)\ , && \Phi_1(x) =
  \half{1} \partial \matq(x) \dot{\matq}(x) \Rightarrow a_1 \delta^2(x-y)\ ,
  \label{eq:quant23.1} \medsp
  & \Psi_0^{\pm \pm} (x) = \half{1} \dot{\matq}^{\pm}(x) \matq^{\pm}(x) \Rightarrow
  b^{\pm \pm}_0
  \delta^2(x-y)\ , && \Psi_1^{\pm \pm}(x) = \half{1} \partial \matq^{\pm} \matq^{\pm} \Rightarrow b^{\pm \pm}_1
  \delta^2(x-y)\ , \label{eq:quant23.2} \medsp
  &\Pi_0^{\pm}(x) = \dot{\matq}(x) \matq^{\pm}(x) \Rightarrow c^\pm_0 \delta^2(x-y)\ . && \label{eq:quant23.3}
\end{align}
From the matter interaction terms in the gauge-fixed Lagrangian \eqref{eq:quant20}
\begin{equation}
  \label{eq:quant24}
  \itindex{L}{int} = 2 i \Phi_0 p_{++} - \sqrt{2} \Psi^{++}_0 p_{++} + \sqrt {2}
  \Psi^{--}_0 p_{--} + 2 i \Pi_0^+ p_+\ ,
\end{equation}
the lowest order vertices can be determined by solving  to first order in
localized matter \eqref{eq:quant23.1}-\eqref{eq:quant23.3} the \emph{classical} equations of motion of
the geometrical variables involved. To this end the
asymptotic integration constants must be chosen in a convenient way. Following
the calculations of the purely bosonic case \cite{Kummer:1998zs,Grumiller:2002dm}
$q^\phi(x^0 \rightarrow \infty) = x^0$, which implies $q^{++}(x^0 \rightarrow
\infty) = i$. In addition $p_{--}(x^0 \rightarrow \infty) = i e^Q$ may now be
imposed. Finally we have to fix the asymptotic value of the Casimir function
\eqref{eq:bosonicC}: $C(x^0 \rightarrow \infty) = C_\infty$. Due to the matter
interactions $\extd C \neq 0$, but the conservation law receives contributions
from the matter fields as well \cite{Bergamin:2003mh}.
The exact form of the fermionic integration constants depends on
$C_\infty$. If $C_\infty = 0$ there exists (asymptotically) a second Casimir
function $\tilde{c}$ \cite{Ertl:2000si,Bergamin:2003am}. Then the value of
$\tilde{c}$ determines one asymptotic integration constant from the
fermions.

The equations of motion of the dilaton, the dilatino and $q^{++}$ can be
solved explicitly for general
asymptotic values $q^+_\infty$ and $q^-_\infty$. However, the dependence of
$q^{--}$ on the different target-space variables $q^I$ is more complicated (cf.\
\eqref{eq:quant17.3}) and the fermion dependent part (soul) of that differential equation cannot be
integrated for general functions $Q$ and $W$. In the current work this
restriction does not have far-reaching consequences: In any scattering problem
the most natural and technically manageable boundary conditions have
asymptotically vanishing values of the fermion fields and thus  all fermionic
integration constants are set to zero: $q^+_\infty = q^-_\infty = 0$. Then the Casimir function has no soul while its body becomes
\begin{equation}
  \label{eq:quant25}
  \begin{split}
    C &= - m_\infty + \Bigl[ i (2 i a_0 - \sqrt{2} b_0^{++}) \Bigl(m_\infty +
    \bigl[ W \bigr]_{y^0} \Bigr) + \sqrt{2} i e^Q b_0^{--}\Bigr] h\ .
  \end{split}
\end{equation}
Square brackets are used to emphasize the terms which depend on $y^0$. The new
abbreviation $h = \theta(y^0-x^0) \delta(x^1 - y^1)$ has been used in this equation. $m_\infty$ is the integration constant
of $q^{--} = i e^{-Q} m_\infty + \ldots$, which, however, turns out to be equivalent to the asymptotic
value $-C_\infty$. Notice that all contributions except this integration
constant are proportional to $h$ and thus all geometric variables may be
replaced by their asymptotic values in that equation. Of course, this is
equivalent to the statement that $C$ is constant in the absence of matter fields. 

\subsection{Four point vertices}

To simplify the solution of the equations of motion for $p_I$, which determine the vertices
according to \eqref{eq:quant24}, we select the special asymptotic geometry
with $C_\infty = 0$. This implies \cite{Bergamin:2003mh} the
BPS condition\footnote{Clearly the asymptotic fields cannot obey both
  supersymmetries. This would imply that $q^{\pm \pm}$ and $p_{\pm \pm}$
  vanish asymptotically and thus the (bosonic) line element would be
  degenerate in this limit.} and for Minkowski ground-state models space-time becomes
asymptotically flat.
To first order in localized matter the relevant equations of motion can be
written as (cf.\ \cite{Bergamin:2003mh}):
\begin{align}
  \partial_0 p_{--} &= -i q^{++} Z p_{--} \label{eq:quant26.0} \medsp
  \partial_0 p_{++} &= i p_\phi -i q^{++} q^{--} e^Q Z \label{eq:quant26.1}
  \medsp
  \partial_0 p_\phi &= -i e^Q \bigl( \bigl(e^{-Q} W' \bigr)' +
  q^{++} q^{--} Z' \bigr) \label{eq:quant26.2} \medsp
  \partial_0 p_+ &= \frac{i}{2 \sqrt{2}} q^- e^{-Q/2} (\sqrt{-W})'' p_{--} +i
  e^{-Q/2} (\sqrt{-W})' p_- \label{eq:quant26.3} \medsp
  \partial_0 p_- &= \frac{-i}{2 \sqrt{2}} q^+ e^{-Q/2} (\sqrt{-W})'' p_{--} - i \half{Z}
  q^{++} p_-
\end{align}
Notice that all terms quadratic in the fermions in \eqref{eq:quant26.1} and
\eqref{eq:quant26.2} vanish as they are second order in localized
matter. The solution of  \eqref{eq:quant26.0} is simply $p_{--} = i e^Q$ and the
expansion to first order in the matter fields yields
\begin{equation}
  \label{eq:eq:quant27.0}
  p_{--} = i e^{Q} \bigl(1 - i Z (2 i a_0 - \sqrt{2} b_0^{++}) (x^0 - y^0)
  h\bigr)\ .
\end{equation}
Using the helpful relation $\partial_0 = -i q^{++} \partial_\phi$ the
equation of motion for $p_+$ can be integrated easily. For later convenience
it is abbreviated as
\begin{align}
  \label{eq:quant27.1}
  p_+ &= - \inv{\sqrt{2}} c^+_0 h V_3 (x,y)\ ,\medsp
  \label{eq:quant27.2}
  V_3 (x,y) &=  \left((\sqrt{-W})' - \Bigl[(\sqrt{-W})' \Bigr]_{y^0} \right)\left((\sqrt{-W}) - \Bigl[(\sqrt{-W}) \Bigr]_{y^0} \right)\ .
\end{align}
As the whole expression is proportional to $c^+_0$ all $\phi$
dependent quantities simply depend on $x^0$.

The equation of motion for $p_{++}$ reads
\begin{equation}
  \label{eq:quant28}
  (q^{++} p_{++}')' = - W'' \bigl(1 +i (2 i a_0 - \sqrt{2}
  b_0^{++})\bigr) + \sqrt{2} i b_0^{--} (e^Q h)' Z\ .
\end{equation}
This yields two analytically different types of contributions to the vertices and thus the
solution is written as
\begin{equation}
  \label{eq:quant28.2}
  p_{++} = iW(x^0) + (2 i a_0 - \sqrt{2} b_0^{++}) h  V_1(x,y) +  \sqrt{2}
  b_0^{--} \bigl( \bigl[ e^Q Z\bigr]_{y^0} (x^0 - y^0) h + V_2(x,y) \bigr)\ .
\end{equation}
The differential equation can be solved explicitly for $V_1$
\begin{equation}
  \label{eq:quant29}
  V_1 = \bigl( W' + \bigl[W'\bigr]_{y^0} \bigr) (x^0 - y^0) - 2 \left( W - \bigl[ W \bigr]_{y^0}\right) \ ,
\end{equation}
while $V_2$ in general cannot be integrated:
\begin{equation}
\label{eq:quant30}
 \partial^2_0 V_2 = e^Q Z^2 h
\end{equation}
It is important to realize that the two integration constants hidden in $V_2$
are determined by the asymptotic value ($x^0\to\infty$) of $p_{++}$ in
(\ref{eq:quant28.2}). Because for this value our boundary conditions yield
$iW(x^0)$ both integration constants have to vanish for $x^0>y^0$, so there is
no ambiguity from homogeneous solutions to (\ref{eq:quant30})
anymore. Exploiting that to leading order $\dot{f}:= df/dx^0=df/d\phi$ the differential equation (\ref{eq:quant30}) can be integrated in the regions $x^0>y^0$, yielding $V_2^> = 0$, and in the region $x^0<y^0$
\begin{equation}
  \label{eq:quant43}
  V_2^< =
  \tilde{A}+\tilde{B}x^0+e^{Q(x^0)}+\int^{x^0}\int^{{x^0}'}e^{Q({x^0}'')}\dot{Z}({x^0}'')d{x^0}''d{x^0}'\ .
\end{equation} 
The matching conditions are given by continuity of $V_2$ and its first derivative at $x^0=y^0$ (note that this is not a requirement imposed arbitrarily but a consequence from the $\theta$-function hidden in $h$):
\begin{align}
  \label{eq:quant44}
  0&=\tilde{A}+\tilde{B}y^0+e^{Q(y^0)}-\int^{y^0}\int^{y'}e^{Q(y'')}\dot{Z}(y'')dy''dy'\,,\\
  \label{eq:quant45}
  0&=\tilde{B}+e^{Q(y^0)}Z(y^0)-\int^{y^0}e^{Q(y')}\dot{Z}(y')dy'\,.
\end{align}
Thus, the full vertex $V_2$ is given by $V_2 = V_2^< h$. As a nontrivial
 example we perform the integrations for
the physically relevant class of models\footnote{E.g.\ in the bosonic part of
  spherically reduced Einstein gravity $ a = (D-3)/(D-2)$. In the CGHS model
  $a=1$ \cite{Grumiller:2002nm}.} with $Z=-a/\phi$ 
\begin{equation}
  \label{eq:744}
  \left.V_2\right|_{Z=-a/\phi} = \frac{a}{a+1}\left((x^0)^{-a}-(y^0)^{-a}+ a (y^0)^{-(a+1)}(x^0-y^0)\right) h\,.
\end{equation}

The different vertices $\mathcal{V}$ now follow straightforwardly. $V_1$ gives rise to
a term of the type $\dot{\matq} \dot{\matq} (x^0) \rightarrow \dot{\matq} \dot{\matq}(y^0)$ from
the first, $\dot{\matq}^+ \matq (x^0) \rightarrow \dot{\matq}^+ \matq(y^0)$
from the second and
the mixed vertex from the first two terms in \eqref{eq:quant24}:
\begin{align}
\label{eq:quant31.1}
  \mathcal{V}\bigl(\dot{\matq} \dot{\matq} (x^0) \rightarrow \dot{\matq} \dot{\matq}(y^0)\bigr) &= -
  4 V_1(x,y) h\ , \medsp
\label{eq:quant31.2}
  \mathcal{V}\bigl( \dot{\matq}^+ \matq^+ (x^0) \rightarrow \dot{\matq}^+
  \matq^+(y^0) \bigr) &=  2 V_1(x,y) h\ , \medsp
\label{eq:quant31.3}
  \mathcal{V}\bigl( \dot{\matq} \dot{\matq} (x^0) \rightarrow \dot{\matq}^+
  \matq^+(y^0) \bigr) &=  - 2
  \sqrt{2} i V_1(x,y) h\ .
\end{align}
Considering the last expression it is important to notice that the two
different contributions from \eqref{eq:quant24} add, as $h(x,y) V_1(x,y)$ is
symmetric in $x$ and $y$.

In a similar way the vertices from $V_2$
follow. Here the contributions from the third term in \eqref{eq:quant24} and
the explicitly integrated part $[e^Q Z]_{y^0}$ cancel and one obtains
\begin{align}
  \label{eq:quant32.1}
   \mathcal{V}\bigl( \dot{\matq} \dot{\matq} (x^0) \rightarrow \dot{\matq}^-
   \matq^- (y^0) \bigr) &= 2
   \sqrt{2} i V_2(x,y)\ , \medsp
  \label{eq:quant32.2}
   \mathcal{V}\bigl( \dot{\matq}^+ \matq^+ (x^0) \rightarrow \dot{\matq}^-
   \matq^- (y^0) \bigr) &=  - 2  V_2(x,y)\ .
\end{align}
Finally, $V_3$ yields vertices with mixed initial and final states:
\begin{equation}
  \label{eq:quant33}
  \mathcal{V}\bigl( \dot{\matq} \matq^+ (x^0) \rightarrow \dot{\matq} \matq^+
  (y^0) \bigr) =
  \frac{-i}{\sqrt{2}} V_3(x,y) h
\end{equation}
Note that the set of vertices is invariant under the exchange $\dot{\matq}
\dot{\matq}\leftrightarrow i\dot{\matq}^+ \matq^+/\sqrt{2}$. Moreover, the
vertices $V_1$ and $V_3$ are conformally invariant, but $V_2$ is not. In
addition, it should be pointed out that all vertices vanish in the coincidence
limit $x^0=y^0$. This is of relevance for the elimination of nonlocal loops:
it seems likely that the arguments in favor of such a cancellation presented
in appendix B.2 of ref. \cite{Grumiller:2003dh} can be extended to the present
case (to visualize a ``nonlocal loop'' we refer to the graphs 2,3,4 and 6 in figure B.6 of that reference).

Another remark concerns the special cases where one or more of the vertices
vanish. Clearly, for non-dynamical dilaton, $Z=0$, no contribution arises from
the non-invariant vertex $V_2$. It is worthwhile mentioning that $V_2$ is
independent of $W$, the ``good'' function in the parlance of
\cite{Grumiller:2003dh} and solely depends on the ``muggy'' one. In contrast,
the invariant vertices solely depend on $W$---which, indeed, is the very
reason for their invariance.

\subsection{Implications for the S-matrix}

The vertices derived in the previous section are the first step to derive S-matrix elements. However, they are the most important one, as from now on the calculation is rather straightforward: one has to introduce asymptotic states, which are very simple for vanishing $C_\infty$, build a corresponding Fock space, insert the asymptotic states in the vertices (\ref{eq:quant31.1}-\ref{eq:quant33}) and perform all the integrations involved. Unfortunately, this method already in the bosonic case \cite{Fischer:2001vz} requires a lot of efforts (see appendix F of \cite{Grumiller:2001ea} for details). Thus, we will restrict ourselves to some of the general features that can be discussed without actually performing all these steps.

For $p_+=0$ at the boundary one obtains for the asymptotic states
\begin{align}
& \partial_0\left(\partial_1\matq-E_1^{--}\partial_0\matq \right) = 0\ , \\
& \partial_1\matq^+ - \partial_0 \left(E_1^{--}\matq^+\right) = 0\ , \\
& \partial_0 \left(E_1^{++}\matq^-\right) = 0\ , \label{eq:mamelina}
\end{align}
with $E_1^{--}=-ip_{++}|_{as}=W|_{as}$ and $E_1^{++}=-ip_{--}|_{as}=e^{Q}|_{as}$. The first two of these equations are conformally invariant because in the present gauge $E_1^{--}$ is invariant while $E_1^{++}$ is not. This matches nicely with the conformal invariance of the vertices $V_1$ and $V_3$ derived in the previous subsection. Thus, contributions from (\ref{eq:quant31.1}-\ref{eq:quant31.3}) and from (\ref{eq:quant33}) to the S-matrix are also conformally invariant. The last equation above implies that $\matq^-$ is not conformally invariant. However, neither is the vertex $V_2$. Thus, it is conceivable that together with the non-invariance of the asymptotic states the noninvariant vertices $V_2$ also yield an invariant contribution to the S-matrix in (\ref{eq:quant32.1}-\ref{eq:quant32.2}). 

Actually, we will show now that the contributions from $V_2$ with external
legs to the S-matrix are not only conformally invariant but they {\em vanish
  identically}: By definition one of the external legs on the left hand side
of \eqref{eq:quant32.1} and \eqref{eq:quant32.2} consists of $\dot{\matq}^-\matq^-$. But since the solution of (\ref{eq:mamelina}) contains a fermionic integration constant $\matq^-_0(x^1)$ and the latter appears quadratically in $\dot{\matq}^-\matq^-$ this term vanishes identically. Similarly, it can be argued that (\ref{eq:quant31.2}), (\ref{eq:quant31.3}) yield no contribution to the S-matrix.
{\em Thus, for the lowest order tree-level S-matrix one has to take into account only (\ref{eq:quant31.1}) and (\ref{eq:quant33}).}\footnote{Nevertheless, the other vertices will be of relevance for calculations of higher order in loops and matter, which is why we presented them in their full glory.}

For constant $W$ (so-called 
``generalized teleparallel'' theories, including rigid supersymmetry with
cosmological constant) these vertices vanish identically. Less trivial is
the special case $W\propto\phi$ (this family of models has been studied in the
bosonic case by Fabbri and Russo \cite{Fabbri:1996bz} and it includes the CGHS
model \cite{Callan:1992rs}): the vertex $V_1$ vanishes, but $V_3$ does
not. Thus, although the CGHS model exhibits the feature of scattering
triviality \cite{Grumiller:2002dm}, its supersymmetric version loses this property.
If $W\propto\phi^2$ the vertex $V_3$ does not contribute. This happens, for
instance, for supergravity extensions of the Jackiw-Teitelboim model
\cite{Barbashov:1979bm}.
In this case the lowest order supergravity scattering amplitude is equivalent
to its bosonic counter part.

\section{Measure and 1-loop action}\label{sec:loops}
In eq.\ \eqref{eq:quant22} the proper definition of the remaining path
integral measure over the matter fields remained open, which is needed in
calculations of matter loops. We follow the steps performed already in the purely
bosonic case  \cite{Grumiller:2002nm,Grumiller:2003mc}: as a suitable
choice of the measure is known for the quantization of the matter fields on a
fixed background \cite{Toms:1987sh}, the path integral in \eqref{eq:quant22} is reduced to that
problem. Nevertheless, the situation is more complicated in the present
application: The result for the supersymmetric path integral measure
\cite{Rocek:1986iz} is derived from a linear realization of supersymmetry
while the gPSM based version deals with a non-linear one. Thus, we suggest to
extend the approach of \cite{Grumiller:2003mc} in such a way that the results
of ref.\ \cite{Rocek:1986iz} should remain applicable to our case.

The path integral measure of the linear theory may be written as $\mathcal{D}
\tilde{M} = \mathcal{D} (E^{1/2} M)$, where $M = f - i \theta \lambda +
\half{1} \theta^2 H$ is the matter
multiplet in superspace and $E$ the superdeterminant. The covariant fields
$\tilde{f}$, $\tilde{\lambda}_\alpha$ and $\tilde{H}$ depend on the
vielbein components $e^a_m$, the (superspace) gravitino\footnote{\label{foot321}The gravitino in
  the MFS model
  $\psi_m^\alpha$ is not identical with the superspace field $\Psi_m^\alpha$, but
  related by $\Psi_m^\alpha=\psi_m^\alpha - \frac{i}{8} Z e_m^a
  \epsilon_{ab} (\chi \gamma^b)^\alpha$ \cite{Bergamin:2003am}.}
$\Psi_m^\alpha$ and the auxiliary fields of the matter sector
$A$. Thus auxiliary background fields $\check{e}^a$, $\check{\Psi}^\alpha$
and $\check{A}$ are introduced, their ``on-shell'' values are determined by means of the
operators
\begin{gather}
  \label{eq:quant51.1}
\begin{align}
  \hat{e}_1^{\pm \pm} &= - i \varfrac{\jp^{\mp \mp}}\ , &  \hat{e}_0^{--} &= -i\ , & \hat{e}_0^{++}
  &= 0\ , \
\end{align}\medsp
\label{quant:51.2}
 \hat{A} = - \half{1}(u' + u Z) - \inv{8} Z'
  \varfrac{\jq_-} \varfrac{\jq_+}\ ,\medsp
\label{eq:quant51.3}
\begin{align}
  \hat{\Psi}_1^\pm &= \mp i \varfrac{\jp^\mp} - \inv{4 \sqrt{2}} Z \varfrac{\jp^{\mp\mp}}\varfrac{\jq_\mp}\ , &
  \hat{\Psi}_0^- &= -\inv{4 \sqrt{2}} Z \varfrac{\jq_+} \ , & \hat{\Psi}_0^+ &= 0\ ,
\end{align}
\end{gather}
where in all dilaton-dependent functions the replacement
$q^\phi \rightarrow -i \varfrac{\jq_\phi}$ must be made. The path integral is
rewritten as
\begin{equation}
  \label{eq:quant52}
  \mathcal{W}[\jq_I, \jp^I,J,J_\alpha] = \int
  \mathcal{D}(\check{e}_m^a,\check{\Psi}_m^\alpha,\check{A}) \delta(\check{e}_m^a-\hat{e}_m^a)
  \delta(\check{\Psi}_m^\alpha - \hat{\Psi}_m^\alpha) \delta(\check{A} -
  \hat{A}) \tilde{\mathcal{W}}\ .
\end{equation}
In $\tilde{\mathcal{W}}$ all geometrical variables in the measure may be replaced by
auxiliary variables and thus the integration reduces to a quantization on a
fixed background.
In our gauge the necessary redefinitions read\footnote{As the calculation of
  this section are performed at the Lagrangian level, canonical coordinates
  $\matq$, $\matq^{\pm}$ are again written as $f$, $\lambda^{\pm}$.}:
\begin{align}
  \label{eq:quant50}
  \tilde{f} &= \sqrt{\check{e}} f & \Tilde{\lambda}^\alpha &= \sqrt{\check{e}} \lambda^\alpha
  + \half{1} \sqrt{\check{e}} f (\check{\Psi}^a \gamma_a)^\alpha & \tilde{H} &= \sqrt{\check{e}} H + \check{\Psi} \tilde{\lambda} +
  \half{1} \tilde{f} (\check{A} - \inv{\check{e}} \check{\Psi}_0^- \check{\Psi}_1^+)
\end{align}
Taking into account the ambiguous terms from \eqref{eq:quant22}
$\tilde{\mathcal{W}}$ in this gauge may
be written as
\begin{multline}
  \label{eq:quant51}
  \tilde{\mathcal{W}}[\check{e}^a_m,\check{\Psi}^\alpha_m,\check{A},\jp^I,\jq_I,J,J_\alpha] =
  \int \mathcal{D}(\tilde{f},\tilde{\lambda}_\alpha,\tilde{H}) \exp\Bigl(i
  \intd{\diff{^2 x}} \bigl(i \check{e}_{++} \dot{f}^2 + \dot{f} (\partial f - 2 i \check{\Psi}_+
  \lambda^+)\medsp - \inv{\sqrt{2}} \check{e}_{++} \dot{\lambda}^+ \lambda^+ + \inv{\sqrt{2}}
  \check{e}_{--} \dot{\lambda}^- \lambda^- - \frac{i}{\sqrt{2}} \lambda^+ \partial \lambda^+ -
  \frac{i}{2} \check{e}_{--} H^2 +
  f J + \lambda^\alpha J_\alpha + \Delta L \bigr) \Bigr)
\end{multline}
Here, $\Delta L$ represents all terms, which do not contribute to the
quadratic part in the matter fields. To make contact with
the calculations from the linear theory, an additional Gaussian factor $\check{e} H^2$
has been introduced (which in the present gauge becomes $\check{e}_{--} H^2$).

After the reformulation of the matter loop expansion as quantization on a
fixed background, existing results from the literature may be taken
over\footnote{Though our formulation of supergravity deals with a non-linear
  realization thereof the matter action \eqref{eq:cmMFS} is---up to the
  Gaussian factor $H^2$---equivalent to the one obtained from superspace. Thus
no additional complications with auxiliary fields arise. However, it should be
mentioned that our argument relies on an important assumption: The results
obtained from the regularized linear theory can be used directly in our
formalism after having integrated out all geometric variables.}. We do not present
any formulae in component expansion, but the more
convenient superspace expressions. Of course, in any concrete applications the
component expansion will become manifest. However it is expected that many
terms thereof will be irrelevant for the leading quantum corrections.

After a careful implementation of the source terms, the supersymmetric
extension of the Polyakov action \cite{Polyakov:1981rd} in a gauge-independent
description\footnote{The superspace conventions are taken from
  \cite{Ertl:2001sj,Bergamin:2003am}. $\mathcal{J} = J_H + i \theta^\alpha
  J_\alpha + \half{1} \theta^2 J$ is the superfield of sources, $S$ the
  supergravity multiplet and $E$ the superdeterminant of ref.\ \cite{Howe:1979ia}.} becomes
\cite{Gieres:1988js}
\begin{equation}
  \label{eq:quant53}
  \Stext{eff} = \int_X \int_Y \Bigl( \inv{24 \pi} (E S)(X) \Delta(X,Y) S(Y) -
  \inv{2} \mathcal{J}(X) \Delta(X,Y) \mathcal{J}(Y) \Bigr)\ .
\end{equation}
Here $X = (x,\theta)$ and $Y = (y,\tilde{\theta})$ are two sets of superspace
coordinates and $\Delta$ the Green function of the quadratic superspace derivative $\half{1} D^\alpha D_\alpha
\Delta = \delta^4(X-Y)$.

\section{Conclusions}\label{sec:conclusions}
A background independent non-perturbative quantization of two dimensional
dilaton supergravity has been presented. The relevant steps rely on the
first order formulation \cite{Bergamin:2003am} of superfield supergravity \cite{Park:1993sd} in terms of a graded
Poisson-Sigma model \cite{Schaller:1994es}, and a Hamiltonian path integral quantization. Remarkably, no
quartic ghost terms arise.

In the matterless case \emph{all} integrations can be done exactly. The ensuing effective action is---up to
boundary terms---equivalent to the gauge-fixed classical action, which
confirms the local quantum triviality of the model as expected from the
bosonic case \cite{Haider:1994cw,Kummer:1997hy,Kummer:1998zs}. Nevertheless,
there exist interesting non-local quantum correlators. For a specific class
thereof a topological interpretation of the non-perturbative quantum effects exists.
It could be of interest to apply these results to super Liouville theory (cf.~e.g.~\cite{Nakayama:2004vk}).

With matter couplings the integration over all geometrical variables still
can be performed exactly, but the matter interactions have to be treated
perturbatively. For this case the (non-local) four-point vertices have been
determined to lowest order in matter. As in the bosonic
theory the phenomenon of virtual black holes
\cite{Grumiller:2000ah,Grumiller:2002dm} is encountered, but supergravity
yields a richer structure of vertices. In particular the supersymmetric CGHS
model---in contrast to its bosonic counterpart---does not exhibit scattering
triviality at tree level. Finally it has been argued that matter loop
corrections should be obtainable from quantization on a
fixed background. In particular, the one loop results follow from the
super-Polyakov action.  

A natural next step will be the calculation of
S-matrix elements for simple scenarios by virtue of the vertices
(\ref{eq:quant31.1})-(\ref{eq:quant31.3}) and (\ref{eq:quant33}), in analogy to the bosonic case \cite{Fischer:2001vz}.
A different aspect to be investigated in detail are matter loops. At least for
specific (simple) models explicit computations of corrections (analogous to the results of
ref.\ \cite{Grumiller:2003mc} in the bosonic theory) should be feasible.

So far, all results on matter interactions are restricted to the minimally
coupled, conformal case.
One might try to extend the quantization procedure to non-minimally coupled
fields and/or self-interactions. However, additional technical difficulties
arise in these cases. Indeed, an important simplification in the calculations
of sections \ref{sec:constraints} and \ref{sec:3} has been the fact that
$G^{++}_{(m)}$ is independent of $p_{--}$, $G^{+}_{(m)}$ independent of $p_-$
etc. These restrictions are lost together with non-minimal coupling or
self-interaction, which makes the calculation of the constraint algebra much
more complicated (cf.\ the difficulties that already arose in the purely bosonic case
\cite{Grumiller:2001ea,Grumiller:2002nm}). Also it is not obvious that the homological perturbation
theory still stops at Yang-Mills level. If this were no longer the case, an
important ingredient justifying the approach outlined in this work may be
lost. Nevertheless, preliminary calculations lead to promising results which
suggest that some generalizations can be treated by the program outlined in
this work.

Finally, two of the present authors have shown recently \cite{Bergamin:2004sr} that the
gPSM approach applies to models with extended
supergravity as well, which suggests further interesting generalizations of the current
work.

\subsection*{Acknowledgement}
The authors are grateful to P.\ van Nieuwenhuizen and D.~V.~Vassilevich for
interesting discussions, especially concerning the 1-loop results and the
fermionic correlators. In addition we thank P.\ van Nieuwenhuizen for
collaboration on the contents of appendix \ref{app:B} during the early stages
of this paper. This work has been supported by projects P-14650-TPH,
P-16030-N08 and J-2330-N08 of the Austrian Science Foundation (FWF). Part of
it has been performed during the workshop ``Gravity in two dimensions'' at the
International Erwin-Schr\"odinger Institute in Vienna.

\appendix
\section{Notations and conventions}
\label{sec:notation}
These conventions are identical to
\cite{Ertl:2000si,Ertl:2001sj}, where additional explanations can be found.

Indices chosen from the Latin alphabet are commuting (lower case) or
generic (upper case), Greek indices are anti-commuting. Holonomic coordinates
are labeled by $M$, $N$, $O$ etc., anholonomic ones by $A$, $B$, $C$ etc., whereas
$I$, $J$, $K$ etc.\ are general indices of the gPSM. The index $\phi$ is used to indicate the dilaton component of
  the gPSM fields:
  \begin{align}
    X^\phi &= \phi & A_\phi &= \omega
  \end{align}

The summation convention is always $NW \rightarrow SE$, e.g.\ for a
fermion $\chi$: $\chi^2 = \chi^\alpha \chi_\alpha$. Our conventions are
arranged in such a way that almost every bosonic expression is transformed
trivially to the graded case when using this summation convention and
replacing commuting indices by general ones. This is possible together with
exterior derivatives acting \emph{from the right}, only. Thus the graded
Leibniz rule is given by
\begin{equation}
  \label{eq:leibniz}
  \mbox{d}\left( AB\right) =A\mbox{d}B+\left( -1\right) ^{B}(\mbox{d}A) B\ .
\end{equation}

In terms of anholonomic indices the metric and the symplectic $2 \times 2$
tensor are defined as
\begin{align}
  \eta_{ab} &= \left( \begin{array}{cc} 1 & 0 \\ 0 & -1
  \end{array} \right)\ , &
  \epsilon_{ab} &= - \epsilon^{ab} = \left( \begin{array}{cc} 0 & 1 \\ -1 & 0
  \end{array} \right)\ , & \epsilon_{\alpha \beta} &= \epsilon^{\alpha \beta} = \left( \begin{array}{cc} 0 & 1 \\ -1 & 0
  \end{array} \right)\ .
\end{align}
The metric in terms of holonomic indices is obtained by $g_{mn} = e_n^b e_m^a
\eta_{ab}$ and for the determinant the standard expression $e = \det e_m^a =
\sqrt{- \det g_{mn}}$ is used. The volume form reads $\epsilon = \half{1}
\epsilon^{ab} e_b \wedge e_a$; by definition $\ast \epsilon = 1$.

The $\gamma$-matrices are used in a chiral representation:
\begin{align}
\label{eq:gammadef}
  {{\gamma^0}_\alpha}^\beta &= \left( \begin{array}{cc} 0 & 1 \\ 1 & 0
  \end{array} \right) & {{\gamma^1}_\alpha}^\beta &= \left( \begin{array}{cc} 0 & 1 \\ -1 & 0
  \end{array} \right) & {{\gthree}_\alpha}^\beta &= {(\gamma^1
    \gamma^0)_\alpha}^\beta = \left( \begin{array}{cc} 1 & 0 \\ 0 & -1
  \end{array} \right)
\end{align}

Covariant derivatives of anholonomic indices with respect to the geometric
variables $e_a = \extd x^m e_{am}$ and $\psi_\alpha = \extd x^m \psi_{\alpha m}$
include the two-dimensional spin-connection one form $\omega^{ab} = \omega
\epsilon^{ab}$. When acting on lower indices the explicit expressions read
($\half{1} \gthree$ is the generator of Lorentz transformations in spinor space):
\begin{align}
\label{eq:A8}
  (D e)_a &= \extd e_a + \omega {\epsilon_a}^b e_b & (D \psi)_\alpha &= \extd
  \psi_\alpha - \half{1} {{\omega \gthree}_\alpha}^\beta \psi_\beta
\end{align}

Light-cone components are very convenient. As we work with spinors in a
chiral representation we can use
\begin{align}
\label{eq:Achi}
  \chi^\alpha &= ( \chi^+, \chi^-)\ , & \chi_\alpha &= \begin{pmatrix} \chi_+ \\
  \chi_- \end{pmatrix}\ .
\end{align}
For Majorana spinors upper and lower chiral components are related by $\chi^+
= \chi_-$, $ \chi^- = - \chi_+$, $\chi^2 = \chi^\alpha \chi_\alpha = 2 \chi_- \chi_+$. Vectors in light-cone coordinates are given by
\begin{align}
\label{eq:A10}
  v^{++} &= \frac{i}{\sqrt{2}} (v^0 + v^1)\ , & v^{--} &= \frac{-i}{\sqrt{2}}
  (v^0 - v^1)\ .
\end{align}
The additional factor $i$ in \eqref{eq:A10} permits a direct identification of the light-cone components with
the components of the spin-tensor $v^{\alpha \beta} = \frac{i}{\sqrt{2}} v^c \gamma_c^{\alpha
  \beta}$. This implies that $\eta_{++|--} = 1$
and $\epsilon_{--|++} = - \epsilon_{++|--} = 1$. The
$\gamma$-matrices in light-cone coordinates become
\begin{align}
\label{eq:gammalc}
  {(\gamma^{++})_\alpha}^\beta &= \sqrt{2} i \left( \begin{array}{cc} 0 & 1 \\ 0 & 0
  \end{array} \right)\ , & {(\gamma^{--})_\alpha}^\beta &= - \sqrt{2} i \left( \begin{array}{cc} 0 & 0 \\ 1 & 0
  \end{array} \right)\ .
\end{align}

\section{Details relevant for the path integral}\label{app:B}

\subsection{Boundary Terms}
\label{sec:3.1}
The action \eqref{eq:gPSMaction} may consistently be extended by a boundary
term
\begin{equation}
  \label{eq:bc1}
  \Stext{gPSM} = \int_M \extd X^I \wedge A_I + \half{1} P^{IJ} A_J \wedge A_I - \int_{\partial M} f(C) X^I A_I
\end{equation}
with $f(C)$ an arbitrary function depending on the Casimir function. Beside
the bulk field equations \eqref{eq:gPSMeom1} and \eqref{eq:gPSMeom2} this yields
the following conditions on the boundary
\begin{align}
\left. f({C})X^I\de A_I \right|_{\partial M} &= 0\ , \label{eq:psmpvn8} 
\\
\left.  \de X^I \left[\left(f(C)-1\right)A_I+ \partial_I C f'(C) 
X^J A_J \right] \right|_{\partial M} &= 0\ . \label{eq:psmpvn9}
\end{align}

\subsubsection{Simplest boundary conditions}

A consistent\footnote{We mean consistency as defined in 
\cite{Lindstrom:2002mc}: boundary conditions arise from (1) extremizing the 
action, (2) invariance of the action under symmetry transformation and (3) 
closure of the set of boundary conditions under symmetry transformations.} way 
is fixing $f({C})=0$ and $\de X^I=0$ at the 
boundary, which for a time-like boundary means $\left.X^I\right|_{\partial M}
= X^I(r)$, where in the gauge chosen in this work $r = r(x^0)$. For the time being we will assume this boundary prescription, which 
has been used e.g. in the quantization of spherically reduced gravity 
\cite{Kummer:1998zs,Grumiller:2000ah}.

With respect to the symmetry variation \eqref{eq:symtrans}
the action (\ref{eq:bc1}) with $f({C})=0$ transforms into a surface integral
\eq{
\de_\eps \Stext{gPSM} = \int_{\partial M} \extd X^I \eps_I\,.
}{eq:psmpvn11}
For the current boundary prescription this term vanishes since $X^I$ are fixed
on the (time-like) boundary. Note that also further variations and/or symmetry 
transformations of (\ref{eq:psmpvn11}) vanish, since $\eps$ depends only on 
the world-sheet coordinates $x$ and the fields $X$, and the symmetry variation
of $X$ again yields a function of $X$. 

The commutator of two symmetry variations
\begin{align}
\left[\de_{\eps_1},\de_{\eps_2}\right]X^I &= \de_{\eps_3} X^I\,, 
\label{eq:psmpvn12} \\
\left[\de_{\eps_1},\de_{\eps_2}\right]A_I &= \de_{\eps_3} A_I + \left(\extd X^J+
P^{JK}A_K\right)\partial_J \partial_I P^{RS} \eps_{1\,S} \eps_{2\,R}\,, 
\label{eq:psmpvn13} 
\end{align}
in general only closes on-shell,
with
\eq{
\eps_{3\, I} = \partial_I P^{JK} \eps_{1\,K} \eps_{2\,J} +
P^{JK} \bigl( \eps_{1\, K} \partial_J \eps_{2\, I} - \eps_{2\, K} \partial_J
\eps_{1\, I} \bigr)\ .
}{eq:psmpvn14}
By applying two consecutive symmetry variations (in order to get insight into
trivial large gauge transformations) we obtain
\eq{
\de_{\eps_1}\de_{\eps_2} L = \int_{\partial M} \left( \extd X^I P^{JK} \eps_{1\,K}
 \partial_J \eps_{2\,I} - P^{IJ}\eps_{1\,J}\extd\eps_{2\,I}\right)\,.
}{eq:psmpvn15}
For fixed $X^I$ at the boundary the first term vanishes, while the second one
can yield only something in the direction orthogonal to $\extd X^K$. E.g. in
purely bosonic first order gravity the boundary values were fixed to $X^k=X^k(r)$ and thus
(in Schwarzschild coordinates) only a $dt$ component can survive in 
(\ref{eq:psmpvn15}). Thus, the boundary action is of the form $\int dt M 
\dot{p}_M$. This is essentially the result that Kuchar obtained for
the Schwarzschild black hole \cite{Kuchar:1994zk}. It is also very similar to 
what Gegenberg, Kunstatter and Strobl obtained in Casimir-Darboux coordinates
(cf. eqs. (38-42) in \cite{Gegenberg:1997de}).
The application of further symmetry transformations does not change the 
structure of this result. 

\subsubsection{Other boundary conditions}

Now we discuss other possible boundary prescriptions, always assuming that
on the boundary $A_I\neq 0 \neq X^I$ in general (because e.g. in gravity 
natural boundary conditions would be most ``unnatural'' since the metric would 
degenerate). Instead of $\de X^I=0$ we can require 
$\de A_I=0$, automatically fulfilling (\ref{eq:psmpvn8}). We now want to avoid
$\de X^I=0$ (which would obey (\ref{eq:psmpvn9}) trivially) and therefore
must require
\eq{
\left. \left[(f({C})-1)A_I+ \partial_I{C} f'({C})X^JA_J \right]
\right|_{\partial_{M}} = 0\,.
}{eq:psmpvn16}
For consistency, also $\de_\eps\de A_I=0$ on the boundary. This yields the
condition
\eq{
\partial_L(\partial_I P^{JK})\eps_K = 0\ ,
}{eq:psmpvn17}
which is satisfied for a linear Poisson tensor (i.e. in the 
Yang-Mills case), since in that case also $\eps=\eps(x)$. Then the well-known
relations for these models are reproduced (cf.\ e.g.\ \cite{Vassilevich:1995we} for the Abelian case). For more general 
theories eq. (\ref{eq:psmpvn17}) would restrict our symmetry transformations
asymptotically, which we excluded in this work.
%
%
Other possible boundary prescriptions are either $\de A_I = A_I = 0$ or
$\de X^I = X^I = 0$ at the boundary with arbitrary $f({C})$. But as 
mentioned before neither of them can be used in (super)gravity.

Since a symmetry transformation mixes the components of $X^I$ a consistent set of
mixed boundary prescriptions does not exist in general. E.g. the requirements
$\de X^1=0$, $A_2=\dots=A_J=0$ and $f=0$ fulfill (\ref{eq:psmpvn8}) and 
(\ref{eq:psmpvn9}), but the symmetry variation of $X^1$ yields $P^{1J}\eps_J$,
which in general will depend on all $X^I$, and a variation thereof need not 
vanish.

In summary we conclude that for the most general gPSM only the boundary 
prescription $\de X^I=0$ and $f({C})=0$ is consistent, while for certain
special cases (essentially Yang-Mills) alternative prescriptions are possible.

One might wonder about consistent boundary conditions for the matter fields as
well. Certainly, this is an interesting question when considering global
objects (solitons). However, the quantization presented in this work treats
the geometrical variables non-perturbatively while the matter fields only can be
included in a perturbative framework. In that case all matter fields
can be assumed to fulfill natural boundary conditions. 

\subsection{Ordering}
\label{sec:3.2}
Generic gPSM gravity (with respect to the Poisson bracket \eqref{eq:canonicalbr}) as well as MFS
minimally coupled to matter (with respect to the Dirac bracket \eqref{eq:diracbr}) is free of
ordering problems if we require hermiticity\footnote{Although it is not
  indicated explicitly, all formulas in this subsection refer to operator expressions.} of the Hamiltonian. To this end we
show the validity of three statements:
\begin{enumerate}
\item Any hermitian operator version of the classical Hamiltonian is automatically Weyl
  ordered.

  As the Hamiltonian is a sum over the constraints \eqref{eq:classicalham},
  while the latter are independent of $\bar{p}_I$, it
  is Weyl ordered if the constraints have that property. For the geometrical
  part the statement follows from their linearity in $p_I$. E.g.\ the $G^I_{(g)}$
  can be written in a hermitian version as
  \begin{equation}
    \label{eq:hermGg}
    {G}^I_{(g)} = \partial q^I + \half{1}\bigl(P^{IJ}(q) p_J + (-1)^{J(I+1)} p_J
    P^{IJ}(q)\bigr)\ ,
  \end{equation}
  which is Weyl ordered since every commutator with $q$ from the right is
  compensated by another commutator from the left. Thus in that part of the
  Hamiltonian no ordering terms can appear.

  The situation in the matter Hamiltonian is almost trivial. As this part of
  $H$ is independent of the target space coordinates $q^I$, the Dirac bracket
  does not lead to complications (cf.\ \eqref{eq:dirbr1}-\eqref{eq:dirbr5}). The
  matter fields do appear at most quadratically and thus Weyl ordering is
  trivial.
\item The commutator $[ {G}^K, {{C}_K}{}^{IJ} ]$ vanishes even at the
  quantum level.

  For the gPSM part one simply notices that this commutator (for any
  ordering prescription, even non-hermitian ones) is given by
  \begin{equation}
    [ {G}_{(g)}^K, {{C}_K}{}^{IJ} ] = - P^{KL} \partial_L \partial_K P^{IJ}
    = 0
  \end{equation}
  due to (anti-)symmetry of the Poisson tensor.

  More involved is the discussion of ${G}^I_{(m)}$. Indeed, it seems that there
  could appear contributions to this commutator, whenever a $\matq^+$ or
  $1/p_{++}$ ($\matq^-$ or $1/p_{--}$) hits
  a $q^{++}$ ($q^{--}$) from the structure functions (cf.\
  \eqref{eq:constralgm1}-\eqref{eq:constralgm4}). Also the term $\propto
  p_\pm$ in ${G}^{\pm \pm}_{(m)}$ is not obviously seen to commute. But it turns out
  that the supergravity restrictions \eqref{eq:sugraconstr1} and
  \eqref{eq:sugraconstr2} resolve all problems:
  \begin{itemize}
  \item All structure functions ${{C}_{++}}{}^{IJ}$ are independent of $q^{++}$ as
    the Poisson tensor is linear in that variable.
  \item All structure functions ${{C}_{++}}{}^{IJ}$ are independent of $q^+$ as
    well, which (for the non-trivial terms) is a consequence of
    \eqref{eq:sugraconstr2}.
  \item All structure functions ${{C}_{+}}{}^{IJ}$ are independent of $q^{++}$,
    which can again be read off from \eqref{eq:sugraconstr1} and
  \eqref{eq:sugraconstr2} without use of the explicit solution
  \eqref{eq:mostgensup}-\eqref{eq:mostgensuplast}. Notice that $\partial_{++}
  {{C}_{+}}{}^{++|--} = 0$ due to the first equation in \eqref{eq:sugraconstr2}:
  As $\partial_+ P^{+|--}$ is independent of $\chi^+$, $\partial_+
  \partial_{++} P^{++|--} = 0$.
  \end{itemize}
  Thus we have shown that the commutator $[ {G}_{(m)}^K, {C}_K{}^{IJ} ] = 0$. In
  fact each individual contribution from the l.h.s.\ of that relation vanishes
  separately.
  \item The commutator of two Weyl ordered constraints yields the structure 
functions times the Weyl ordered constraints (or, equivalently, the Weyl 
ordered product of constraints times the structure functions).

  Using the result for ${G}_{(m)}^I$ from the second point above this statement is obvious for
  that part of the constraints. For ${G}_{(g)}^I$ we can use their form
  \eqref{eq:hermGg}. The commutator between ${G}^I$ and ${G}^J$ yields several terms, 
including ordering terms proportional to $\delta(0)$. By rearranging the terms
such that the structure functions are on the left side and using the Jacobi
identity \eqref{eq:nijenhuis} we obtain
\begin{equation}
\label{eq:quantconstralg}
\left[{G}^I,{G}^J\right]={G}^K {C}_K{}^{IJ} = (-1)^{K(I+J+1)}
{C}_K{}^{IJ} {G}^K\ ,
\end{equation}
without any ordering terms. The second equality in \eqref{eq:quantconstralg} holds due to point 2.
\end{enumerate}

\providecommand{\href}[2]{#2}\begingroup\raggedright\endgroup

\end{document}